\documentclass{IEEEoj}
\usepackage{cite}
\usepackage{amsmath,amssymb,amsfonts}
\usepackage{algorithmic}
\usepackage{graphicx,color}
\usepackage{textcomp}

\usepackage{algorithm}
\usepackage{array}
\usepackage{cite}
\usepackage{picinpar}
\usepackage{amsmath}
\usepackage{url}
\usepackage{flushend}
\usepackage[latin1]{inputenc}
\usepackage{colortbl}
\usepackage{soul}
\usepackage{multirow}
\usepackage{pifont}
\usepackage{alltt}
\usepackage[hidelinks]{hyperref}
\usepackage{enumerate}
\usepackage{siunitx}
\usepackage{breakurl}
\usepackage{epstopdf}
\usepackage{pbox}
\usepackage{booktabs}
\usepackage{amssymb}

\def\BibTeX{{\rm B\kern-.05em{\sc i\kern-.025em b}\kern-.08em
    T\kern-.1667em\lower.7ex\hbox{E}\kern-.125emX}}
\AtBeginDocument{\definecolor{ojcolor}{cmyk}{0.93,0.59,0.15,0.02}}
\begin{document}
% \receiveddate{XX Month, XXXX}
% \reviseddate{XX Month, XXXX}
% \accepteddate{XX Month, XXXX}
% \publisheddate{XX Month, XXXX}
% \currentdate{27 March, 2024}
% \doiinfo{OJCOMS.2024.032700}

\title{Innovative RIS Prototyping Enhancing Wireless Communication with Real-Time Spot Beam Tracking and OAM Wavefront Manipulation}

\author{Yufei Zhao\IEEEauthorrefmark{1}, Yuan Feng\IEEEauthorrefmark{2}, Afkar Mohamed Ismail\IEEEauthorrefmark{1},
Ziyue Wang\IEEEauthorrefmark{1},
Yong Liang Guan\IEEEauthorrefmark{1}\IEEEmembership{(Senior Member, IEEE)},
Yongxin Guo\IEEEauthorrefmark{2}
\IEEEmembership{(Fellow, IEEE)},
and Chau Yuen\IEEEauthorrefmark{1}
\IEEEmembership{(Fellow, IEEE)}}
\affil{School of Electrical and Electronic Engineering, Nanyang Technological University, 639798, Singapore}
\affil{Department of Electrical and Computer Engineering, National University of Singapore, 117583, Singapore}
\corresp{CORRESPONDING AUTHOR: Yuan Feng (e-mail: y.feng@nus.edu.sg).}
\authornote{This work was partly supported by the National Research Foundation, Singapore, and Infocomm Media Development Authority under its Future Communications Research \& Development Programme, Grant No. FCP-NTU-RG-2022-011 and Grant No. FCP-NTU-RG-2022-020, partly by Temasek Laboratories@NTU seed research project No. TLSP23-13.}
\markboth{Innovative RIS Prototyping Enhancing Wireless Communication with Real-Time Spot Beam Tracking and OAM Wavefront Manipulation}{Yufei Zhao \textit{et al.}}

\begin{abstract}
Reconfigurable metasurface, also known as Reconfigurable Intelligent Surfaces (RIS), with its flexible beamforming, low-cost, and easy industrial deployment characteristics, presents many interesting solutions in wireless application scenarios. This paper presents a sophisticated reconfigurable metasurface architecture that introduces an advanced concept of flexible full-array space-time wavefront manipulation with enhanced dynamic capabilities. The practical 2-bit phase-shifting unit cell on the RIS is distinguished by its ability to maintain four stable phase states, each with ${90^ \circ }$ differences, and features an insertion loss of less than 0.6 dB across a bandwidth of 200 MHz. All reconfigurable units are equipped with meticulously designed control circuits, governed by an intelligent core composed of multiple Micro-Controller Units (MCUs), enabling rapid control response across the entire RIS array. Owing to the capability of each unit cell on the metasurface to independently switch states, the entire RIS is not limited to controlling general beams with specific directional patterns, but also generates beams with more complex structures, including multi-focus 3D spot beams and vortex beams. This development substantially broadens its applicability across various industrial wireless transmission scenarios. Moreover, by leveraging the rapid-respond space-time coding and the full-array independent programmability of the RIS prototyping operating at 10.7 GHz, we have demonstrated that: 1) The implementation of 3D spot beams scanning facilitates dynamic beam tracking and real-time communication under the indoor near-field scenario; 2) The rapid wavefront rotation of vortex beams enables precise modulation of signals within the Doppler domain, showcasing an innovative approach to wireless signal manipulation; 3) The beam steering experiments for blocking users under outdoor far-field scenarios, verifying the beamforming capability of the RIS board.
\end{abstract}


\maketitle

\section{INTRODUCTION}
\IEEEPARstart{I}{n} today's urban landscapes, the dense layout of buildings presents substantial obstacles to wireless services, especially affecting vehicles and Unmanned Aerial Vehicles (UAVs) due to potential signal blockages that deteriorate communication quality and may even cause disruptions \cite{open1,open2}. Traditional approaches to maintaining consistent connectivity struggle in areas with tall structures, narrow passages, and various barriers. Recently, reconfigurable metasurfaces, or Reconfigurable Intelligent Surfaces (RISs), have emerged as a viable solution, providing unparalleled control over the propagation of Electro-Magnetic (EM) waves \cite{open3,open5}. These surfaces are notable for their adaptable beamforming abilities, affordability, and straightforward integration into industrial settings, marking significant progress in meeting the sophisticated needs of wireless services \cite{cui1,long1,dai1,R2-2}.

While traditional metasurfaces show great promise, their practical use faces several challenges due to complex design and control system limitations, such as fixed states, low-order phase quantization, low response speed, and cumbersome control logic. Previous work has initiated a wave of research into intelligent metasurfaces, such as, \cite{Access} and \cite{Yang} pioneered practical RIS demonstration prototyping in 2020 with limited 1-bit phase modulation capability per unit, which may result in 3 dB energy loss in beamforming caused by the low bit quantization \cite{bit1,bit2}. A year later, \cite{chengqiang} unveiled a 2-bit RIS system, which is limited to row-by-row unit control, restricting beam scanning to just one dimension. The design suggested by \cite{control}, involving a single Micro-controller Unit (MCU) and several shift registers, did offer a way to independently control multiple states per unit cell. However, the limited number of I/O ports on the MCU meant that updating all shift registers took too much time, slowing down the RIS's response speed, and limiting its effectiveness in space-time-coding situations.
%This delay significantly impacts the system's ability to quickly track fast-moving objects like UAVs, 
% Despite the considerable potential of conventional metasurfaces, their applications have been hampered by a series of functional limitations. These constraints, primarily stemming from the complexity of array design and control systems, include fixed operational states, low-order phase quantization, and inefficient row-by-row control mechanisms. For instance, references \cite{Access} and \cite{Yang} pioneered practical RIS demonstration prototyping in 2020 with limited 1-bit phase modulation capability per unit, which may result in 3 dB energy loss in beamforming \cite{bit1,bit2}. In 2021, reference \cite{chengqiang} introduced a 2-bit RIS experimental system, which was constrained by its ability to control units only on a row-by-row basis, thereby limiting beam scanning to a single spatial dimension. Authors in \cite{control} proposed a design integrating a single Microcontroller Unit (MCU) and multiple shift registers, thereby enabling independent control of multiple states for each unit cell. Due to the limited I/O ports of the single MCU, updating the states of all shift registers requires a considerable amount of time, consequently affecting the overall response speed of the RIS. This delay will undermine the dynamic response capabilities, making it less suitable for rapid beam tracking of high-speed moving targets, e.g., UAVs \cite{Ding}.

Drawing inspiration from these remarkable contributions above, we identified the essential need for a RIS system capable of quickly and independently controlling each unit without compromising dynamic responsiveness, especially critical in situations requiring fast adaptability. Our research introduces an innovative 2-bit phase-quantized RIS design that enables independent control of each unit while effectively addressing challenges related to space-time dynamic response. This system utilizes multiple MCUs to facilitate swift and efficient state changes, essential for fast beam tracking and space-time coding. Our 2-bit phase-quantized approach and hierarchical control of unit cells significantly improve beamforming capabilities, allowing for the precise and fast adjustment of cone beams and the creation of complex beam structures like spot and vortex beams. This advancement enhances wireless services flexibility and offers solutions tailored to the demanding conditions of modern environments.

Additionally, addressing the complexities of wireless near-field scenarios, we have evolved the traditional 2D beam direction scanning into 3D spatial spot beam scanning, and introduced a spot beamforming phase compensation codebook scheme for the phase-quantized RIS. Furthermore, a 3D spot beam tracking algorithm specifically designed for rapidly moving users has been developed. Through feedback between the user and the control system, stable spot beam locking can be achieved, significantly bolstering the RIS prototyping's adaptability to complex EM environments. 
% Then, the experimental validations of this system, as demonstrated in vortex rotational Doppler modulation and highly dynamic communication experiments for mobile users, add a practical dimension to the theoretical advancements.
% It effortlessly handles diverse and complex beamforming requirements, as validated through full-wave EM simulation analyses.
From theory to practice, innovations at multiple levels in both software and hardware have endowed our RIS prototyping with robust adaptability to various applications. In different experimental scenarios, we demonstrate that (a) 3D spot beam scanning enables dynamic tracking, guaranteeing stable and real-time data transmission with the receiver; (b) vortex beams' space-time wavefront manipulation facilitates precise signal modulation in the Doppler domain, offering a novel signal processing method; (c) beams steering for blocking users under outdoor far-field scenarios, verifying the beamforming capability of the RIS board.
% that (a) the energy coverage of communication signals can be adaptively enhanced by our RIS prototyping; 

To clarify, the main contributions of this research can be summarized as follows:

\begin{enumerate}
\item Firstly, this paper proposes an innovative 2-bit RIS unit cell design through a non-resonant method, that achieves four stable and reconfigurable phase states, while maintaining consistent amplitude attenuation (below 0.6 dB) within the effective bandwidth (over 200 MHz). Incorporating the full-array independent control flexibility, this design furnishes an efficient solution tailored to fulfill the intricate beam steering and space-time-coding requirements.
\item Next, in terms of programmable hardware support, we designed a RIS signal control circuit board consisting of multiple MCUs, enabling \textit{parallel} and \textit{pre-configured} phase states controlling for each independent Radio Frequency (RF) unit cell. The RIS prototyping constructed based on this multi-MCU based hierarchical control system not only achieves highly dynamic beam tracking in horizontal, vertical, and depth (near and far) dimensions, but also enables more complex structured EM waves, e.g., rotational vortex beams. This opens up richer possibilities for future RIS-based wireless transmission solutions. 
\item Moreover, this study addresses the characteristics of the EM near-field in multiple reflection scenarios and tailors a spot beamforming codebook scheme for RIS-aided wireless communications, enhancing the precision of 3D spatial beam tracking. A direct and efficient method for implementing spot beam tracking has been proposed in this paper. It introduces a fast spot beam scanning algorithm, facilitating rapid and robust tracking of moving users through comprehensive scans of predetermined spatial regions, which enables stable data transmission within complex reflection interference.
\item Finally, from theory to practice, innovations at multiple levels in both software and hardware have endowed our RIS prototyping with robust adaptability to various applications. It effortlessly handles diverse and complex beamforming requirements, as validated through full-wave EM simulation analyses. In various experimental scenarios, we demonstrated the powerful space-time wavefront manipulation capabilities of the RIS prototype, providing an efficient testing platform for a wide range of future wireless applications.
% (a) the implementation of 3D spot beam scanning allows for dynamic beam tracking, ensuring stable and real-time data transmission to mobile users. the energy coverage of communication signals can be adaptively enhanced by our RIS prototyping; (b) the rapid wavefront rotation of vortex beams enables precise modulation of signals within the Doppler domain, introducing an innovative approach to signal processing; (c) , e.g., UAV.
% from theory to practice, our RIS prototype exhibits extensive adaptability through innovations in both software and hardware. In diverse experiments, we demonstrate that: (a) 3D spot beam scanning enables dynamic tracking, guaranteeing stable, real-time communication with mobile users; (b) vortex beams' rapid wavefront rotation facilitates precise signal modulation in the Doppler domain, offering a novel signal processing method.
\end{enumerate}

The rest of this paper is organized as follows: Section II delves into the 2-bit phase-quantized RIS implementation and its dynamic control system, with a wide comparison of other related works. Section III explains the RIS codebook design, and introduces an efficient 3D spot beam tracking approach, which is verified by the real-time tracking communication experiment. Section IV showcases the vortex wavefront high dynamic manipulation approach, offering a novel signal processing method in the Doppler domain. In Section V, an outdoor far-field test has been implemented to verify the beamforming capability of our RF RIS board. In the end, Section VI concludes this paper and discusses the implications of the findings, potential applications, and avenues for future research.

\begin{table*}[htbp]
\caption{Comparisons of the reflective RIS prototyping from implementations to applications}
\begin{center}
\begin{tabular}{c|c|c|c|c|c|c|c|c|c}
\toprule%\hline
\textbf{Ref.} & \multicolumn{6}{c|}{\textbf{Unit cell of the RIS board}} & \multicolumn{2}{c|}{\textbf{Control Circuitry}} & \textbf{Functions or} \\
\cmidrule{2-9}%\cline{2-7}
\textbf{works} & Components & Center Freq. & Polar.$^{\mathrm{a}}$ & States & Ins. loss & Bandwidth & Controller & Control DoF$^{\mathrm{b}}$ & \textbf{Applications}\\
\midrule%\hline
\cite{long1} & 4 PINs & 3.5 GHz & LP, CP & 2 bit & 2 dB & 180 MHz & FPGA & By units & Polar. converter\\
\midrule%\hline
\cite{Access} & 1 PIN & 11.5 GHz & LP & 1 bit & 1 dB & 1 GHz & MCU, FPGA & by units & Beam steerer\\
\midrule%\hline
\cite{chengqiang} & 2 PINs & 9.5 GHz & LP & 2 bit & 1 dB & 100 MHz & FPGA & By columns & Space-time-coding \\
\midrule%\hline
\cite{control} & 1 PIN & 60.25 GHz & LP & 1 bit & 5.3 dB & 450 MHz & FPGA & By units & Beam steerer \\
\midrule%\hline
\cite{resonance} & 2 PIN & 8.0 GHz & LP & 2 bit & 1 dB & 500 MHz & - & By columns & AM PM tunable \\
\midrule%\hline
\cite{Deng} & 4 PIN & 3.6 GHz & LP (daul) & 2 bit & $<$1 dB & 400 MHz & FPGA & By rows & Beam steerer \\
% \midrule%\hline
% \cite{EuCAP} & 1 varactor & 24.5 GHz & LP & 2 bit & <3 dB & 400 MHz & - & - & Beam steerer \\
\midrule%\hline
\cite{R1-2} & 1 PIN & 26.0 GHz & LP & 1 bit & $<$1 dB & 400 MHz & FPGA & - & Beam steerer \\
\midrule%\hline
\cite{R2-4} & 5 PIN & 2.3 GHz & LP, CP & 2 bit, non- & 1 dB & 400 MHz & FPGA & By units & Beam steerer \\
\midrule%\hline
\cite{R1-1} & RF Switch & 5.3 GHz & LP & 3 bit, non- & - & 118 MHz & MCU with & By units & Beam steerer \\
 &  &  &  & resonant &  &  & serial bus &  &  \\
\midrule%\hline
\cite{R3-4} & 1 PIN & 28.5 GHz & LP (daul) & 1 bit & 2 dB & 2 GHz & FPGA & By units & Beam steerer \\
\midrule%\hline
\cite{R4-4} & 1 PIN & 27.5 GHz & LP & 1 bit & $<$3 dB & 5.25 GHz & - & By units & Beam steerer \\
\midrule%\hline
\cite{R5-4} & 1 Varactor & 3.6 GHz & LP & 2 bit & 1.2 dB & $<$400 MHz & FPGA & By columns & Time-digi.-coding \\
\midrule%\hline
\cite{R7-4} & 4 Varactors & 3.7 GHz & LP & Multi-bit & $<$3 dB & - & FPGA & By columns & Time-digi.-coding \\
\midrule%\hline
This & 3 PINs & 10.7 GHz & LP & 2 bit, non- & 0.6 dB & \>200 MHz & Hierarchical & By units & Spot beam steerer, \\
work &  &  &  & resonant &  &  & MCUs &  & rotational vortex \\

\bottomrule%\hline
%${l_3} = -2$ & -74 dBm & -75 dBm & -56 dBm & -79 dBm \\
%\hline
%${l_4} = +2$ & -78 dBm & -76 dBm & -76 dBm & -57 dBm \\
%\hline
%\multicolumn{5}{1}
\end{tabular}
\label{tab_com}
\end{center}
{$^{\mathrm{a}}$ LP: Linear polarization; CP: Circular polarization. $^{\mathrm{b}}$ DoF: Degree of freedom. It indicates the degree of flexibility with which the entire RIS array can be controlled. It depends on the specific applications of each RIS prototyping.}
\end{table*}

\section{2-bit RIS Design with Its Powerful Control System}
\subsection{Phase-Quantized Unit Cell}
\begin{figure}[htbp]
\centering
\includegraphics[width=3.4in]{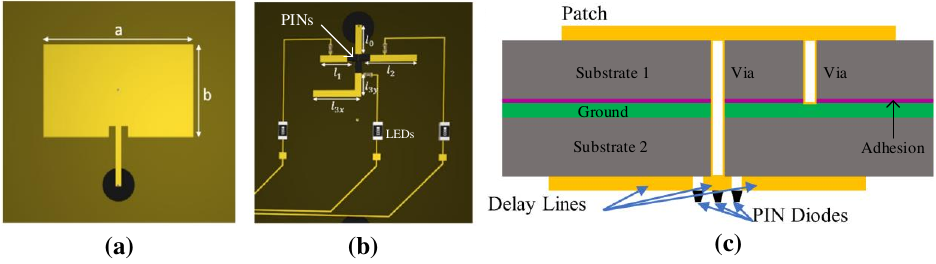}
\caption{Illustrates the innovative design of the RIS unit cell with respect to (a) front-side, (b) back-side, and (c) laminated construction. ($a=13.8~\rm{mm}$, $b=8.3~\rm{mm}$, ${l_0}=3.46~\rm{mm}$, ${l_1}=3.84~\rm{mm}$, ${l_2}=5.64~\rm{mm}$, ${l_3}=7.64~\rm{mm}$, width is $0.6~\rm{mm}$ for each line)}
\label{fig1}
\end{figure}

In RIS-aided wireless solutions, the critical aspect for achieving reconfigurability lies in how to design each independent RF discrete unit. Conventional RIS utilizes RF units designed on resonant principles, enabling control of EM wave phase and amplitude through adjustable electronic components that manipulate the equivalent circuit's impedance and reactance to adjust the resonant frequency \cite{nano}. However, as the number of reconfigurable states increases, the complexity of the unit cell design escalates \cite{resonance,proceeding}. Moreover, nonlinear coupling between different states significantly impacts the independence and stability of state transitions within the unit cell. These problems pose significant challenges for both the design and practical implementation of the whole RIS prototyping \cite{arxiv}. To address these limitations, we introduce a novel non-resonance 2-bit phase-quantized reconfigurable unit cell in this paper, which has been illustrated in Fig. \ref{fig1}. 
To avoid inter-state coupling effects during state transitions of the unit cell \cite{resonance}, we employ transmission line theory to devise a reconfigurable RF non-resonance structure \cite{theory}, capable of applying four distinct discrete phase changes to incident EM waves while maintaining consistency in the amplitude.
%meticulously engineered to modulate the phase of incoming EM waves

As shown in Fig. \ref{fig1}, each unit cell consists of three metallic and three dielectric layers. The topmost metallic layer features a rectangular copper patch with dimensions of 13.8 mm $\times$ 8.3 mm. This patch serves as the initial point of engagement with the EM wave, capturing it for further interaction with the other RIS components. Mounted on two 1 mm-thick Taconic TLX-8 dielectric substrates (Substrates 1 and 2). Sandwiched between these substrates is another adhesion layer composed of Rogers RO4450F. Along with the copper ground layer beneath it, which has a thickness of 0.035 mm. A delay line, referred to as ${l_0}$, is connected to the patch through a via. 

The transmission line lengths in unit cell design are computed based on the principle of achieving specific phase shifts for the reflected signals. The goal is to control the phase of the reflected waves to achieve desired beamforming patterns. The phase shift $\Delta \phi $ introduced by a transmission line of length $L$ is given by $\Delta \phi {\rm{ = }}\varepsilon L$, where $\varepsilon $ is the phase constant of the transmission line, defined as $\varepsilon  = {{2\pi } \mathord{\left/{\vphantom {{2\pi }{{\lambda _{\rm{c}}}}}} \right. \kern-\nulldelimiterspace} {{\lambda _{\rm{c}}}}}$, with ${{\lambda _{\rm{c}}}}$ being the wavelength of the operating frequency on the high-frequency laminates. For a 2-bit RIS, we need to achieve four distinct phase states ($0$, $\pi/2$, $\pi$, and $3\pi/2$). The required length $L$ for each phase shift is computed using the following formula as,
\begin{equation} 
L = {{\Delta \phi } \mathord{\left/
 {\vphantom {{\Delta \phi } \varepsilon }} \right.
 \kern-\nulldelimiterspace} \varepsilon } = {{\Delta \phi  \cdot {\lambda _{\rm{c}}}} \mathord{\left/
 {\vphantom {{\Delta \phi  \cdot {\lambda _{\rm{c}}}} {2\pi }}} \right.
 \kern-\nulldelimiterspace} {2\pi }}.
\end{equation}
Specifically, for each phase shift: 
\begin{equation} 
\left\{ {\begin{array}{*{20}{l}}
{{L_0} = 0 + \delta ,}&{{\rm{for~0~phase~shift,}}}\\
{{L_1} = {{{\lambda _{\rm{c}}}} \mathord{\left/
 {\vphantom {{{\lambda _{\rm{c}}}} 4}} \right.
 \kern-\nulldelimiterspace} 4} + \delta ,}&{{\rm{for~}}\pi {\rm{/2~phase~shift,}}}\\
{{L_2} = {{{\lambda _{\rm{c}}}} \mathord{\left/
 {\vphantom {{{\lambda _{\rm{c}}}} 2}} \right.
 \kern-\nulldelimiterspace} 2} + \delta ,}&{{\rm{for~}}\pi {\rm{~phase~shift,}}}\\
{{L_3} = 3{{{\lambda _{\rm{c}}}} \mathord{\left/
 {\vphantom {{{\lambda _{\rm{c}}}} 4}} \right.
 \kern-\nulldelimiterspace} 4} + \delta ,}&{{\rm{for~3}}\pi /2{\rm{~phase~shift,}}}
\end{array}} \right.
\end{equation}
where $\delta $ is the constant initial length. 
After the initial design phase, the transmission lines are verified using electromagnetic simulation tools such as CST to ensure that the intended phase shifts are accurately achieved. During the physical implementation progress, adjustments are made as necessary to accommodate practical factors such as fabrication tolerances and the properties of the substrate material. 

\begin{figure}[htbp]
\centering
\includegraphics[width=3.4in]{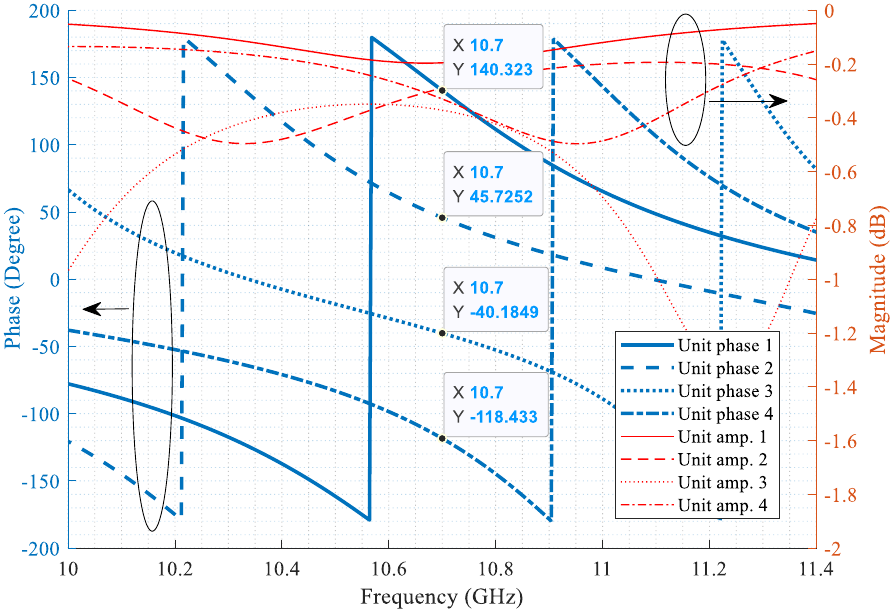}
\caption{Phase and amplitude simulations of the 2-bit unit cell.}
\label{fig2}
\end{figure}

As shown in \ref{fig1}, the bottom layer of each unit houses three additional delay lines, ${l_1}$, ${l_2}$, and ${l_3}$, with lengths of 3.84 mm, 5.64 mm, and 7.64 mm, respectively. Each of these delay lines is connected to delay line ${l_0}$ through individual PIN diodes. These PIN diodes are independently managed by separate signal control lines, each featuring an inductor to isolate Alternating Current (AC) RF signals and mitigate potential interference to the Direct Current (DC) circuit. Additionally, each control line is equipped with an LED to visually indicate the current operational state of the patch. A via grounds the center of the rectangular patch, completing the whole electrical circuit. Engineered to function at a frequency of 10.7 GHz with over 200 MHz bandwidth, the unit cell is capable of four distinct operational states by changing the ON/OFF states of the PINs. The ON/OFF states switching of the PIN diodes forces EM waves to traverse varying lengths of propagation paths within the unit structure, thereby imposing different phase delays on the reflected signal. We simulate the unit cell together with PIN diodes SMP1320 under CST Studio Suite environment, and the results have been shown in Fig. \ref{fig2}. Obviously, the amplitude attenuations of the reflected signal of the four states of each unit remain below 0.6 dB within a 200 MHz bandwidth ($10.6 \sim 10.8$ GHz), while the phase differences are stabilized at about $90^\circ$ among them. It is worth noting that for a 2-bit RIS design, achieving amplitude attenuation within 1 dB is already considered outstanding performance \cite{chengqiang,Deng,EuCAP}. 

As we know, in recent years, research on RIS has flourished. To better showcase the works of various researchers on reflective RIS prototypes over the past few years, a comparative Table \ref{tab_com} has been created that lists the key features of our proposed design alongside those of existing works. As shown in Table \ref{tab_com}, current RIS prototype research varies significantly, primarily due to specialized designs aimed at different application scenarios. There is no standardized design scheme at present. To match actual application scenarios, RIS prototypes focus not only on hardware metasurface design but also on comprehensive system engineering, which includes selecting control chips, circuit design, software algorithms, and communication protocols among various components. For instance, the project in \cite{R1-2} demonstrates an RIS-enabled ultra-massive MIMO system in a practical testbed, and addresses practical challenges by integrating commercial NR base stations and off-the-shelf user equipment. In \cite{R2-4}, Dai \textit{et al.} also successfully achieve 2-bit phase shifting using non-resonant unit cells, which contributes to the industrialization of RIS prototyping. Then, in \cite{R1-1}, Rossanese \textit{et al.} proposed another non-resonant RIS unit cell design based on the RF switch. 
The researchers in \cite{R3-4} and \cite{R4-4} explore azimuth and elevation beam control with independent control mechanisms and 1-bit reconfigurable unit cells. Moreover, References \cite{R5-4} and \cite{R7-4} initially introduced the time-domain digital-coding concept for metasurfaces, which significantly enhanced the flexibility of RIS prototypes and opened up a new research area for future RIS applications. 

Inspired by previous research, in this work, we are dedicated to introducing a novel design concept for the RIS prototype. Firstly, We propose a more straightforward unit cell design through the non-resonant principle to achieve 2-bit phase shifting. In our design, different phase shifts correspond to distinct delay line paths that are independent of each other. When a PIN diode switches to select a different path, it does not affect the other paths, thereby minimizing the interaction between different phase states. The switching speed of the PIN diodes is typically in the nanosecond range, which is sufficient to ensure rapid transitions between phase states. It is worth noting that this approach avoids the repeated adjustments of the equivalent circuit resonance states of unit cells required by many existing works, significantly simplifying the RIS unit cell design process. 

Secondly, as shown in Table \ref{tab_com}, numerous studies have managed to achieve multi-bit state reconfigurability for RIS. However, to transition RIS from a laboratory setting to commercial use, the control of RIS should not be overlooked. Developing a mechanism to achieve more flexible and rapid dynamic responses, enabling the RIS prototype to adapt to more complex application scenarios, remains a challenging issue. Hence, the following sections of this paper introduce a hierarchical control mechanism based on parallel processing and pre-configuration. The control architecture employs multiple MCUs, each equipped with its own storage and computational capabilities. Each follower MCU is responsible for controlling a small subset of RIS unit cells and operates under the directives of a leader MCU. Unlike simple registers, each MCU in our control board has independent storage and computational resources, enabling it to pre-configure various codebook states for the unit cells under its control. Upon receiving instructions from the leader MCU, each follower MCU can rapidly switch states, which significantly accelerates the transition process.

Furthermore, in our design, each unit of the RIS can independently change its response state under the control system, allowing for functionalities such as rapid beam steering, near-field spot beam tracking, and vortex wavefront rotational modulation. All these features will be sequentially demonstrated in the following sections. Additionally, thanks to its flexible space-time digital coding ability, our system can quickly achieve continuous rotational speed shifts of complex vortex beam wavefront. This capability enables the receiver to observe the effects of carrier Doppler frequency shifts in the frequency domain. These are the distinctive features that set the RIS prototyping and the design methodology proposed in this paper apart from previous works.

\subsection{Rapidly-Respond Control Circuitry with Multi-MCU}
\begin{figure}[htbp]
\centering
\includegraphics[width=3.4in]{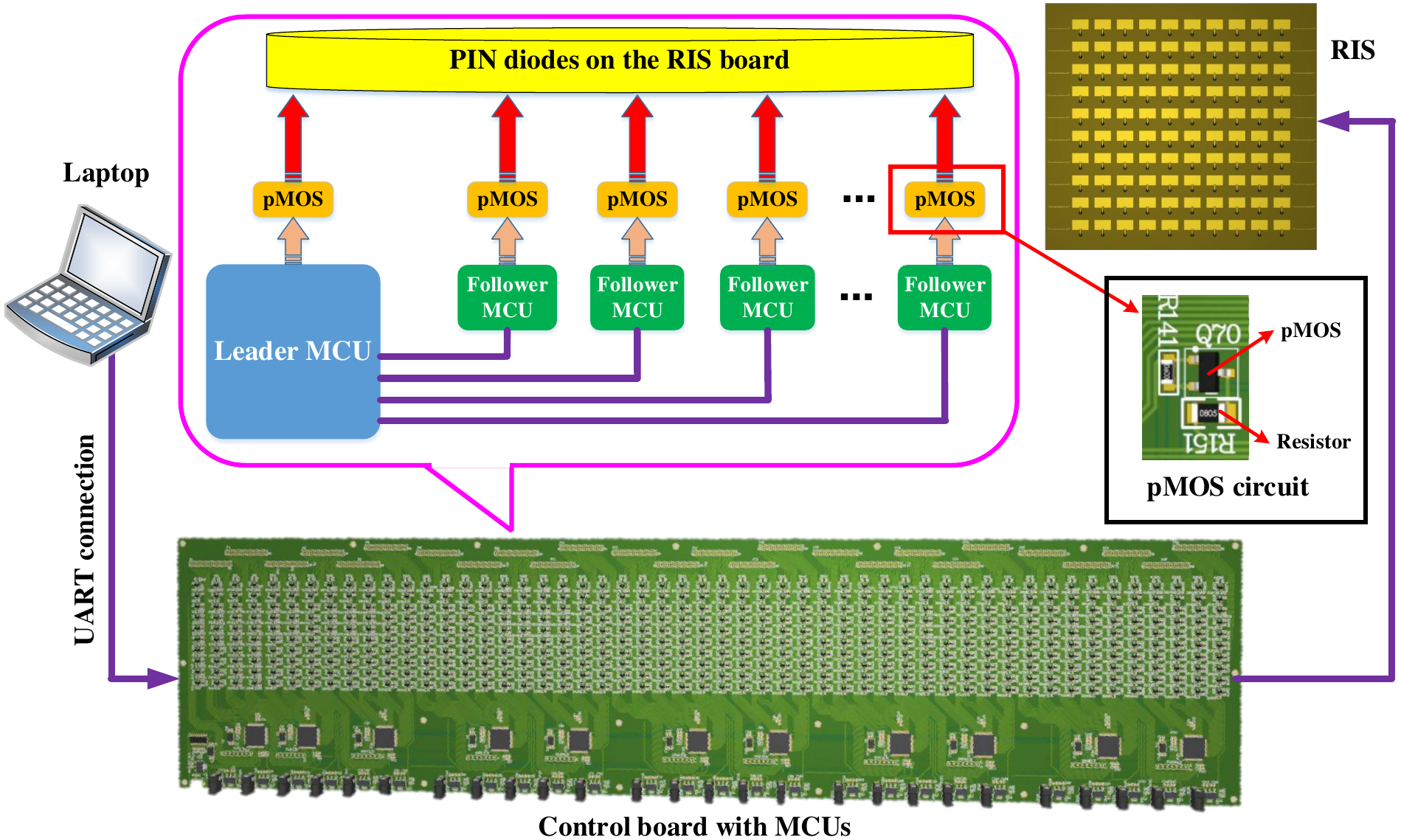}
\caption{Demonstrating the connection relationship and signal transmitting progress of the whole RIS system with control circuitry.}
\label{fig3}
\end{figure}

Specifically, the current RIS prototyping contains 10 $\times$ 10 RF units with an amount of 300 PIN diodes on it, which are connected to a user-defined core signal control board employing multi-MCU chips on it \cite{control}. The dynamic reconfigurable functionality relies on a meticulously designed control board, employing 11 MSP430F5529 MCUs, where one serves as a leader MCU and the remaining 10 functions as follower MCUs. This configuration exemplifies distributed processing, facilitating efficient control of the whole RIS system. As demonstrated in Fig. \ref{fig3}, a total of 520 output pins are formed by soldering the Metal-Oxide-Semiconductor Field Effect Transistor (MOSFET) circuit circuit to each General Purpose Input/Output (GPIO) pin of the MCUs. Specifically, the main MCU has 20 GPIOs left available for controlling the PIN diodes, and each of the 10 follower MCUs has 50 GPIOs available for use. This arrangement provides 520 output pins, creating a sophisticated control mechanism that can handle a multitude of patterns. The MOSFET circuit, consisting of 1 p-channel-MOS (pMOS) and 2 resistors, triggers on negative voltage values, which inversion of input-output logic, wherein '0' represents 0V and '1' represents 3.3V, increases the robustness of the control circuit. While the control board possesses 520 output pins, only 300 are used, connecting to the current RIS board's control lines. (Actually, we over-design the control circuitry for other potential functions in the future. As the level of industrial completion improves, the whole control system will also become more compact and lightweight.)
%Power is supplied to the control board through 21 power lines carrying 5V. A linear voltage regulator then steps down this input to 3.3V, which is further distributed across the board.

Suppose that a codebook ${{\bf{U}}^{10 \times 10}} \in \left\{ {0,1,2,3} \right\}$ can be used to represent the state of each unit cell on the RIS board. Four numbers represent four quantized phase states. Each unit cell is controlled by three PIN diodes, with different ON/OFF state combinations governing the 2-bit phase transition switch. Hence, from this codebook, the binary state of each of the 300 pins can be determined. Binary data is then converted to hexadecimal data, ready to be transmitted to the MCUs.
%in hexadecimal format
The leader MCU, upon power-up, establishes a connection with a PC via USB, employing a Universal Asynchronous Receiver/Transmitter (UART) for data transmission. The MCUs will receive control instructions together with the codebooks of the RIS phase patterns from the PC.
The leader MCU validates the receiving data, extracts relevant information, and prepares it for each follower MCU. The data transmission protocol used is Serial Peripheral Interface (SPI), promoting high-speed data exchange.
Moreover, for the control system, the codebooks are pre-loaded into the leader MCU, which is connected to the other follower MCUs via SPI. Each follower MCU independently controls a small subset of PIN diodes' ON/OFF states. Upon receiving commands, all follower MCUs can operate in parallel without interfering with each other. This parallel processing and pre-configuration mechanism significantly enhances the operational efficiency of the RIS control system, thus increasing the refresh rate of the entire RIS array.

For instance, the current control system has 9 output ports connected to the RIS PIN diodes, with each port's data size being 8 bits (1 byte). Assuming that command data is sent from the leader MCU to the other 10 followers within one clock signal, and each follower receives 9 bytes of data, the total time taken for SPI communication is calculated based on the number of bits sent and the baud rate (in bits per second). Therefore, for 10 followers, 9 bytes per follower, and a baud rate of 1 Mbps, the total action time for SPI communication can be calculated as ${{10 \times 9 \times 8{\kern 1pt} {\kern 1pt} {\rm{bits}}} \mathord{\left/ {\vphantom {{10 \times 9 \times 8{\kern 1pt} {\kern 1pt} {\rm{bits}}} {1{\kern 1pt} {\kern 1pt} {\rm{Mbps}}}}} \right. \kern-\nulldelimiterspace} {1{\kern 1pt} {\kern 1pt} {\rm{Mbps}}}} = $ 0.72 ms. In other words, the control system needs at least 0.72 ms to switch to a phase distribution corresponding to a new codebook, i.e., ${D_{\rm{t}}} = 0.72$ ms which will be described in Sec. \ref{tracking}. Assuming we divide the half-space served by the RIS into $P = 20$ regions of different sizes, and ${t_{\rm{d}}} = 1$ ms, then the time required for the RIS to traverse all codebooks is approximately 15.4 ms for our current hardware design. 
% That means, during this time period, if the mobile user only moves 0.2 meters, it can be calculated that the corresponding speed has reached 47 km/h. The above calculations demonstrate that the control system of the RIS prototyping proposed in this work can meet the mobility requirements of the majority of mobile users in most urban areas, e.g., UAVs, motorized vehicles, or electric bicycles. 
Indeed, the RIS prototype proposed in this paper is still in the initial stages of development, and many signal processing algorithms are not fully embedded in the MCUs but are instead executed using LabVIEW and MATLAB, which leads to noticeable delays in the demos. As the level of industrial completion improves, the speed of time-domain coding processing in this RIS prototype, which has parallel processing and pre-configuration capabilities, will undoubtedly accelerate.

\subsection{Simulation, Fabrication and Measurement}
\begin{figure}[htbp]
\centering
\includegraphics[width=3.0in]{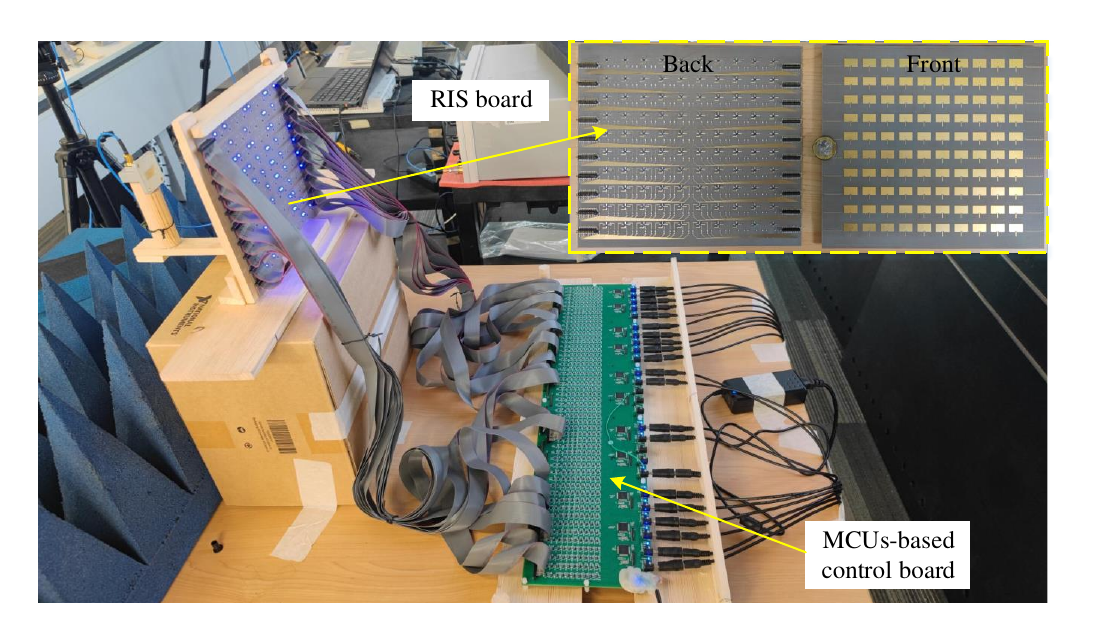}
\caption{Fabrication of the RIS prototyping.}
\label{RIS}
\end{figure}

After fabrication and assembly, the entirety of the RIS along with its control system is depicted in Fig. \ref{RIS}. The RIS panel itself measures 228 mm by 208 mm and comprises 100 unit cells arranged in a 10 $\times$ 10 grid, in addition to 20 plug headers that serve to connect the control lines of each unit cell to the pMOS circuit on the control board. When there is a change in the phase state of the RIS, LEDs positioned on the backside flash in accordance with the ON/OFF status of the PIN diodes, offering a direct method to monitor the operational state of the entire RIS prototyping visually. 

As we know, mutual coupling arises from the EM interaction between closely spaced unit cells, leading to induced currents that affect the current distribution on each unit cell. The altered current distribution due to mutual coupling can result in phase errors and amplitude variations across the array, which negatively impact beamforming. Increasing the spacing between unit cells can significantly reduce coupling effects, but excessively large spacing can lead to higher side lobes in the radiation pattern due to spatial under-sampling. In our design, we carefully chose the unit-cell size to be approximately $3/5$ of a wavelength. This dimension is optimized through CST under periodic boundary conditions. The slightly larger spacing helps to mitigate mutual coupling without significantly affecting the beamforming performance. For the inter-state coupling, our design utilizes microstrip transmission lines of varying lengths to achieve phase shifts. The reconfigurable phase state of each unit cell is directly correlated with the physical length of each transmission line, which minimizes inter-state coupling by ensuring that each phase state corresponds to a unique and independent transmission path. 

\begin{figure}[htbp]
\centering
\includegraphics[width=2.9in]{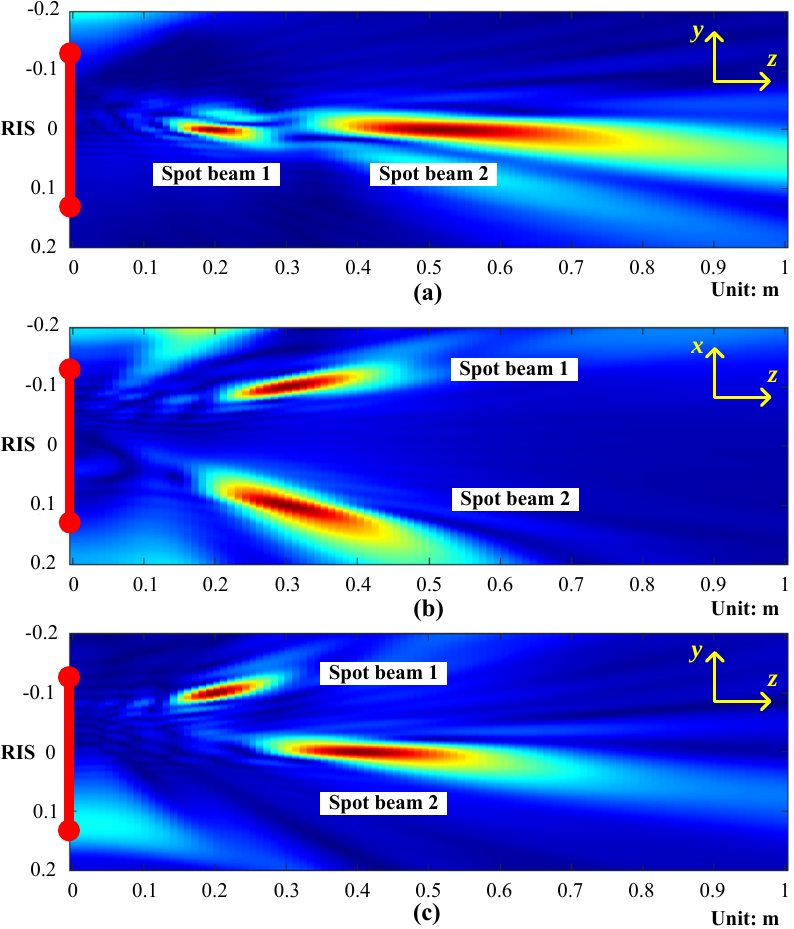}
\caption{Simulation results for the 3D spot beamforming. (a) 2 spot beams along the normal direction of the RIS. (b) 2 spot beams located on either side of the RIS normal direction. (c) 2 spot beams positioned at asymmetric locations in the near field of the RIS.}
\label{fig5}
\end{figure}

% Simulations using MATLAB and CST Studio Suite confirmed the RIS prototyping's beamforming abilities. Figure \ref{fig5}(a) demonstrates how the RIS modulates a transmitter's beam into two distinct energy-dense regions, allowing for simultaneous service to two mobile users without interference. These examples, with spot beams on either side of the RIS's normal (Fig. \ref{fig5}(b)) or at asymmetric positions (Fig. \ref{fig5}(c)), highlight the RIS's capability for multiple spot beamformings via different codebooks. Such beam placement flexibility illustrates the RIS's potential for varied wireless applications like multi-focus power transfer, sensor networks, and direct communications between devices without interference.
Furthermore, the simulations were designed to verify the beamforming capabilities of the RIS prototyping. As illustrated in Fig. \ref{fig5}(a), the beam emanating from the transmitter, once modulated by the RIS, creates two energy-dense regions. These regions materialize concurrently in front of the RIS normal with different locations, demonstrating the RIS's ability to serve two distinct users simultaneously without spatial interference. This scenario, depicted as a representative example calculated via MATLAB, showcases the potential for generating multiple spot beamformings through the RIS by employing various codebooks, further exemplified in Fig. \ref{fig5}(b) and Fig. \ref{fig5}(c). In Fig. \ref{fig5}(b), the two spot beams are positioned on either side of the RIS's normal, while in Fig. \ref{fig5}(c), the spot beams materialize at asymmetric locations in space, predetermined by the system's configuration. This versatility in beam placement underscores the RIS's advanced beamforming potential, enabling precise and flexible targeting to meet diverse wireless applications, e.g., multi-focus wireless power transfer, wireless sensor networks, and device-to-device communications. 

\begin{figure}[htbp]
\centering
\includegraphics[width=3.1in]{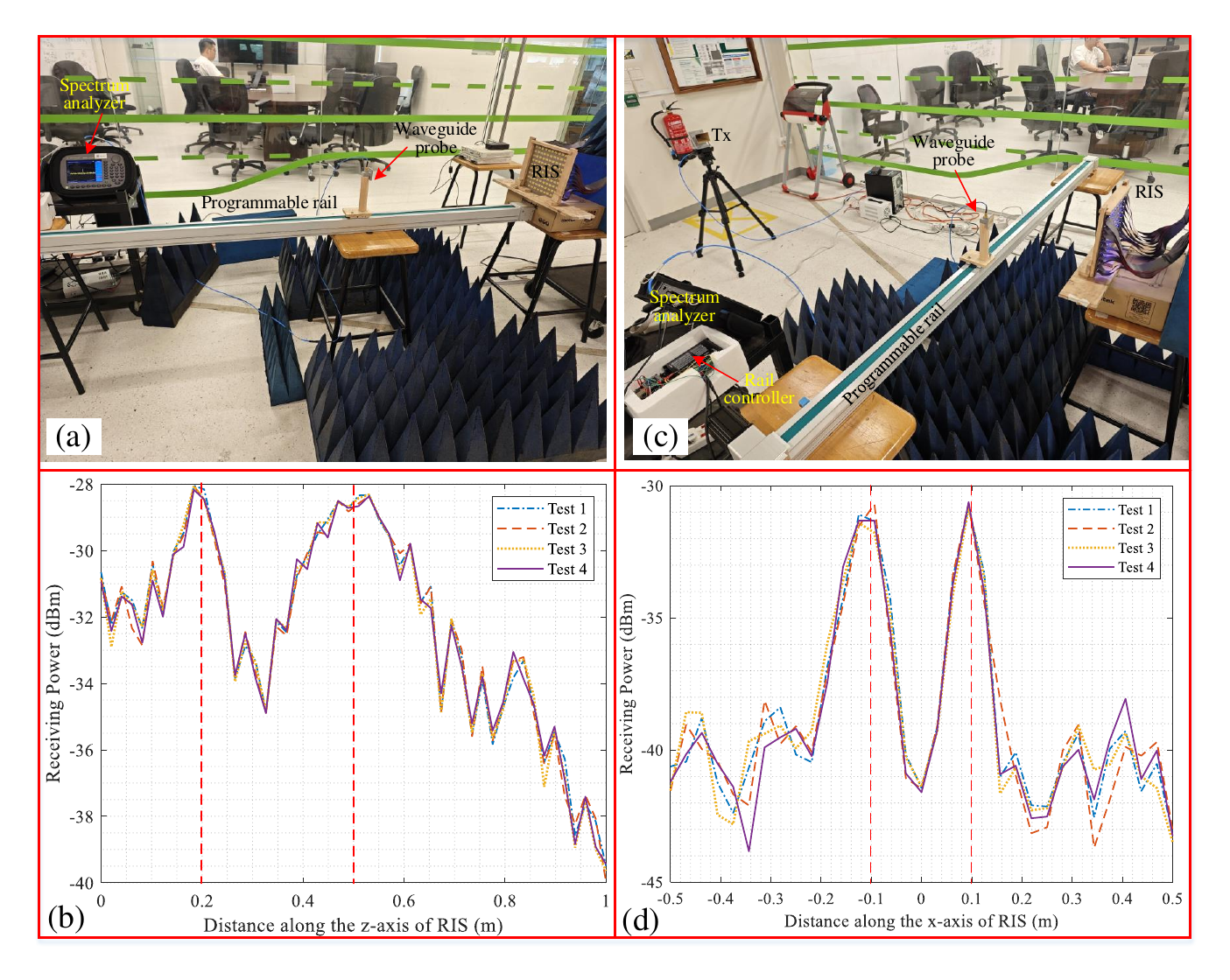}
\caption{Measurement for the 3D spot beamforming. (a) Measurement along the $z$-axis perpendicular to the RIS plane. (b) Energy distribution along the $z$-axis, results from (a). (c) Measurement along the $x$-axis parallel to the RIS plane. (b) Energy distribution along the $x$-axis (0.3 m apart), results from (c). (The red vertical lines indicate the measured positions of the peak power)}
\label{rail}
\end{figure}

Additionally, a measurement experiment is set up to replicate the spot beamforming results observed in the above simulations. As shown in Fig. \ref{rail}, we positioned a rail in front of the RIS. On the rail, there is a slider that can move step-by-step along the rail according to the programming. A waveguide probe mounted on the slider is connected to a spectrum analyzer to measure and record the spatial EM signal energy at each location. The experimental setup is designed to validate the simulation results shown in Fig. \ref{fig5}(a) and Fig. \ref{fig5}(b), corresponding to two spot beamforming schemes, i.e., one perpendicular to the RIS plane and the other parallel to the RIS plane. In the measurement, the same RIS phase distribution pattern was programmed into the MCUs-based control board. As illustrated in Fig. \ref{rail}(a), the linear rail was aligned perpendicular to the RIS plane, and the waveguide probe traversed from near to far along the rail, with the spectrum analyzer capturing the spatial energy distribution. The corresponding measurements, presented in Fig. \ref{rail}(b), consistently revealed two prominent energy peaks at distances of approximately 0.2m and 0.5m from the RIS plane, demonstrating focused beam energy at these points, which is consistent with the simulation results in Fig. \ref{fig5}(a). Similarly, to validate Fig. \ref{fig5}(b), the linear rail was set parallel to the RIS plane at a distance of about 0.3m, and the waveguide probe was programmed to traverse from right to left. The spectrum analyzer systematically recorded the energy distribution along this path, as depicted in Fig. \ref{rail}(c). The findings, summarized in Fig. \ref{rail}(d), displayed consistent dual energy peaks flanking the normal to the center of the RIS, confirming the simulation results in Fig. \ref{fig5}(b).

\begin{figure}[htbp]
\centering
\includegraphics[width=3.0in]{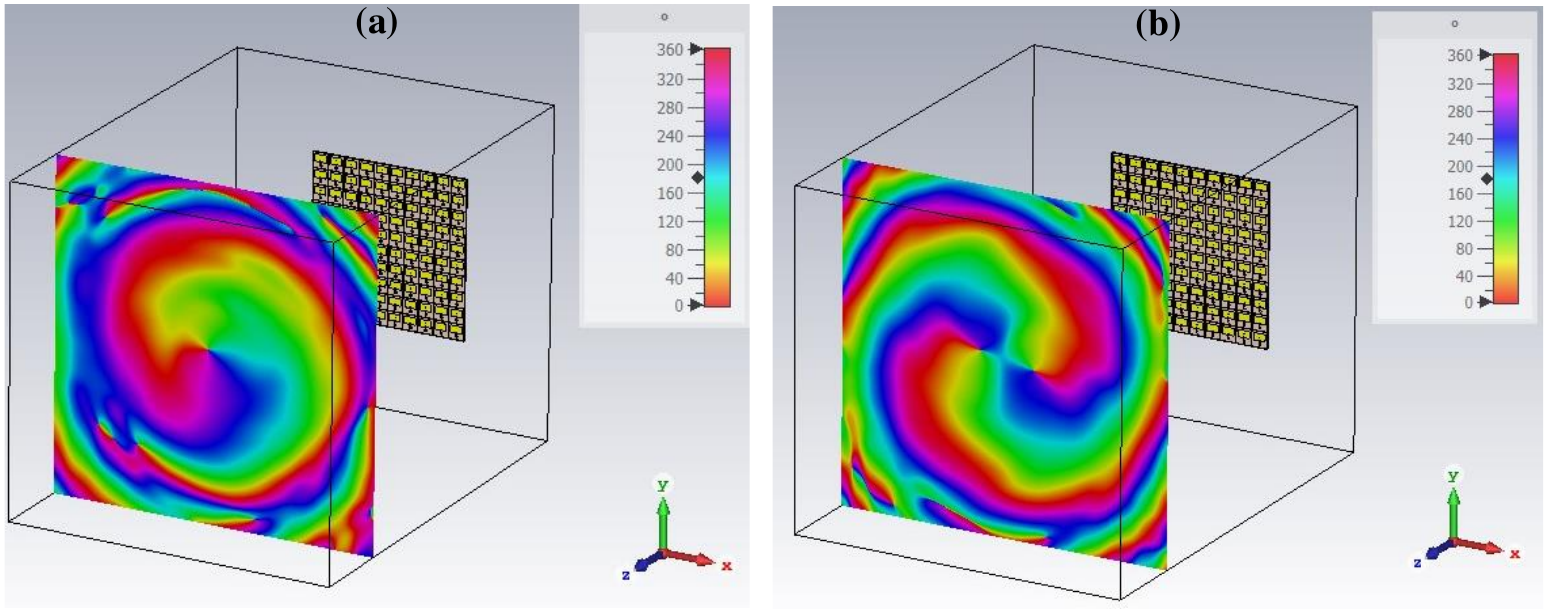}
\caption{Generation for vortex beams under CST Studio Suite. (a) Vortex topology mode 1. (b) Vortex topology mode 2.}
\label{vortex}
\end{figure}

Furthermore, it is worth noting that for some reconfigurable metasurface schemes, although each unit cell has the capability of 2-bit phase states control, only the entire row or column of unit states can be adjusted simultaneously due to limitations in the control system \cite{APMC}. Such RIS typically only possesses the ability for cone beamforming and cannot form spot beams, let alone structured EM waves such as vortex beams. Structured EM waves, characterized by the intricate phase structure within the beam, typically necessitate RIS to have at least simultaneous 2D control capabilities for generation. In this paper, the proposed RIS prototyping allows each unit cell to be independently controlled, with less coupling between different reconfigurable phase states. Fig. \ref{vortex}(a)(b) display two types of full-wave simulation results generating vortex beams in the CST environment, further demonstrating the flexible beamforming capabilities of the proposed RIS prototyping. These results not only highlight the advanced functionality but also its potential to significantly impact the development of many wireless solutions by facilitating more complex and efficient signal distribution strategies, which will be demonstrated in our experiments.

% \begin{table}[htbp]
% \caption{Some main simulation parameters.}
% \begin{center}
% \begin{tabular}{c|c|c}
% \toprule%\hline
% \textbf{Parameter} & \textbf{Value} & \textbf{Dimension} \\
%  \midrule%\hline
%  Central carrier frequency & 10.7 & GHz \\
%  RIS size $M \times N$ & $10 \times 10$ & - \\
%  Size of each unit cell & $18.8 \times 18.65$ & mm \\
%  Distance between Tx and RIS  & 4.0 & m \\
% % Gain of the Tx antenna & 23 & dB \\
%  RF Gain of Tx/Rx & 0 & dB \\
% % Amplifier noise figure & 6.5 & dB \\
% % Amplifier gain & 26.1 & dB \\
%  %Airy lens diameter & 0.6 & m \\
%  Vortex topology modes & +1, -1 & - \\
% % Baseband modulation & QPSK/16-QAM & - \\
%  \bottomrule%\hline
% \end{tabular}
% \end{center}
% \label{parameter}
% \end{table}

\section{RIS-based Spot Beamforming and Tracking Scheme for Rapidly-Respond Solutions} \label{tracking}
\subsection{Codebook Design for RIS Spot Beamforming}
\begin{figure}[htbp]
\centering
\includegraphics[width=3.4in]{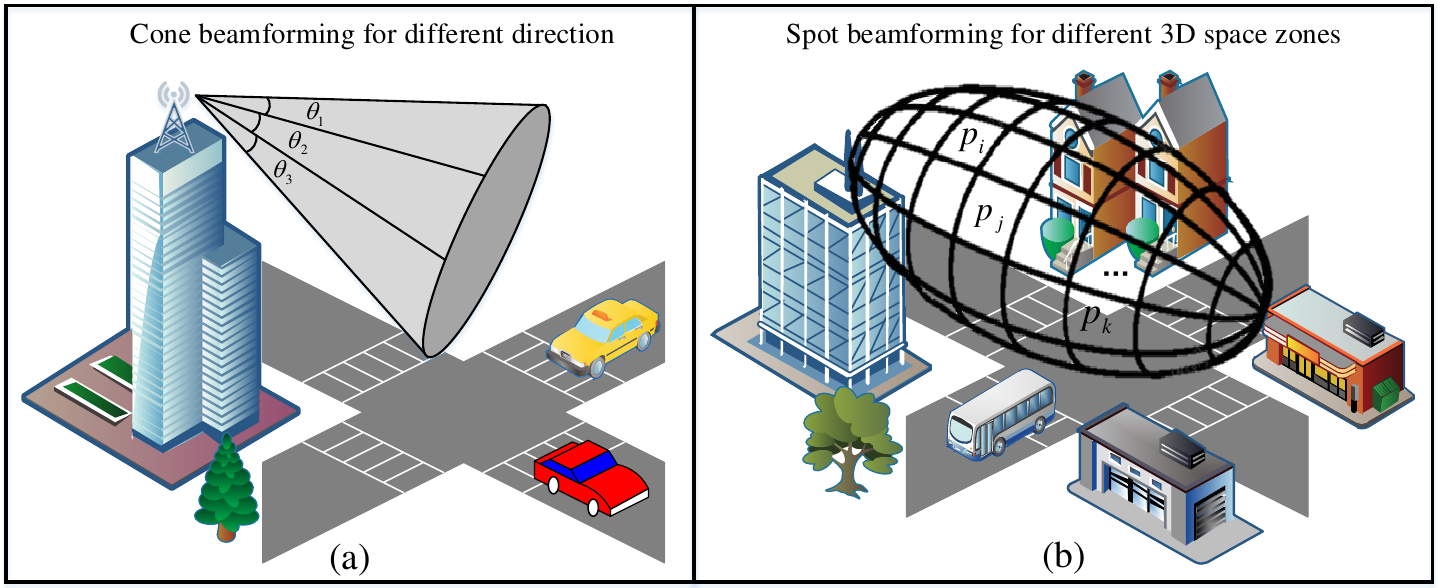}
\caption{2D directional beamforming vs. 3D spot beamforming.}
\label{fig4}
\end{figure}

In general, we consider ${{2{D^2}} \mathord{\left/{\vphantom {{2{D^2}} \lambda }} \right.\kern-\nulldelimiterspace} \lambda }$ as the boundary distinguishing between the far-field and near-field of the RF transmitter, where $D$ denotes the aperture size of the radiator, $\lambda$ is the wavelength. However, in the RIS-assisted urban wireless communication environment, with the increase in the 2D aperture size of the RIS board and the higher carrier frequencies, the assumption of ideal far-field communication conditions no longer holds in practical scenarios \cite{Dai2,Liu1}. This influence is particularly evident in the beam-steering aspect of the antenna array.
As shown in Fig. \ref{fig4}(a), under the traditional far-field assumptions, the beams controlled by the transmitter are cone-shaped beams, i.e., the receiving users are at specific angular directions of the transmitter. In some specific scenarios, such as serving far-field users with a 2D directional beamforming requirement, the unit cells distributed in a particular column or row on the RIS might exhibit an approximately identical phase distribution. This could simplify the control circuit of the RIS. Nonetheless, for comprehensive beam steering at both azimuth and elevation angles, the phase profile requires individual configuration of each unit cell, which also demands a more complex control mechanism \cite{R3-4,R4-4}. Generally, in real-world applications, especially in congested urban environments, both the transmitter and receiver are in the near-field region of the metasurface \cite{Zeng}. In 2023, Sadeghian \textit{et al.} discusses the impact of RIS in indoor environments and suggests that a significantly large RIS might be necessary to make a noticeable difference compared to ambient propagation \cite{R1-3}. In this scenario, what is incident on the RIS board is no longer a planar wave but a spherical wavefront with a structured phase. As demonstrated in Fig. \ref{fig4}(b), beamforming is affected not only by the angular dimension but also by the distance. It has evolved from traditional 2D beam scanning to 3D spatial beam focusing. In this context, distinct from traditional cone beamforming, the new RIS-based codebook design can be referred to as spot beamforming.

% Fig. \ref{fig4}(a) illustrates that under traditional far-field assumptions, transmitter-controlled beams are cone-shaped, targeting users in specific angular directions. In such cases, the RIS's unit cells in a given column or row share nearly the same phase distribution, allowing for a simpler control circuit for far-field applications. However, in dense urban settings, both transmitter and receiver often fall within the metasurface's near-field region, as noted by \cite{Dai1}. This changes the wave incident on the RIS from planar to spherical with structured phases, as shown in Fig. \ref{fig4}(b), making beamforming depend on both angle and distance and shifting from 2D scanning to 3D spatial focusing. This necessitates a shift in RIS codebook design from traditional directional beamforming to spot beamforming for precise targeting.

In the Cartesian coordinate system, assume that the center of the RIS as the coordinate origin, the coordinates of each unit cell on the RIS board can be represented by ${{\bf{u}}_{m,n}} = \left( {{x_{m,n}},{y_{m,n}},{z_{m,n}}} \right)$, $m \in \left\{ {1,2, \ldots ,M} \right\}$, $n \in \left\{ {1,2, \ldots ,N} \right\}$, $M \times N$ is the total number of the discrete unit cells on the RIS board, and the transmitter's coordinates relative to the RIS are denoted as ${{\bf{u}}_{\rm{T}}} = \left( {{x_{\rm{T}}},{y_{\rm{T}}},{z_{\rm{T}}}} \right)$. Then, the half-space served by the RIS can be partitioned into $P$ distinct receiving regions based on varying angles and distances. We can assume the coordinates of the center of each receiving region as ${{\bf{p}}_i} = \left( {{x_i},{y_i},{z_i}} \right),{\kern 1pt} {\kern 1pt} i = 1, \ldots ,P$. Evidently, the number of codebooks recorded by the RIS is directly associated with the granularity of spatial partitioning. Each codebook essentially corresponds to a 2-bit quantized phase distribution state on the RIS board.

Without loss of generality, assume that the RIS is placed at the $x$-$y$ plane. Generally, the positions of the transmitter and RIS are relatively fixed. Based on pre-defined coordinate positions, it is straightforward to calculate the geometric distance from the transmitter to each unit on the RIS as,
\begin{equation} \label{eq1}
{d_{m,n}} = \sqrt {{{\left( {{x_{\rm{T}}} - {x_{m,n}}} \right)}^2} + {{\left( {{y_{\rm{T}}} - {y_{m,n}}} \right)}^2} + {{\left( {{z_{\rm{T}}} - {z_{m,n}}} \right)}^2}}.
\end{equation}
Then, the electric field reaching the $m$-th row and $n$-th column of the RIS can be calculated as,
\begin{equation} \label{eq2}
\vec E\left( {m,n} \right) = \frac{{G\lambda }}{{4\pi {d_{m,n}}}}\exp \left( { - jk{d_{m,n}}} \right),
\end{equation}
where $G$ is the propagation constant associated with physical factors such as antenna gain and radiation pattern, ${{k = 2\pi } \mathord{\left/ {\vphantom {{k = 2\pi } \lambda }} \right. \kern-\nulldelimiterspace} \lambda }$ is the wave vector, $j$ is the imaginary unit. According to the principle of wave interference, after reflection from all units on the RIS, the EM waves reaching a specific spatial grid should coherently superimpose with the same phase \cite{yufei-IoT}. Let's simplify by considering the center point of the grid as the reference for beam interference. In this case, the propagation path difference from distinct units to the reference point can be calculated as,
\begin{equation} \label{eq3}
{f_{m,n,i}} = \sqrt {{{\left( {{x_{m,n}} - {x_i}} \right)}^2} + {{\left( {{y_{m,n}} - {y_i}} \right)}^2} + {{\left( {{z_i} - {z_{m,n}}} \right)}^2}}  - {z_i}.
\end{equation}

If we view the propagation of EM waves as a series of rays, in order to compensate for the phase differences introduced by the spatial positions of distinct unit cells during the propagation process, for each unit cell, the required interference compensation phase offset can be got from the following equation,
\begin{equation} \label{eq4}
{T_{m,n,i}} = \frac{{{{\left( {4\pi } \right)}^2}{d_{m,n}}{d_{m,n,i}}}}{{G{\lambda ^2}}}\exp \left( {jk{d_{m,n}} + jk{f_{m,n,i}}} \right),
\end{equation}
where, ${d_{m,n,i}} = \left| {{{\bf{u}}_{m,n}} - {{\bf{p}}_i}} \right|$. If the RIS simultaneously serves multiple users in different spatial grids, then the compensation on each unit cell is the vector sum of the calculated results from \eqref{eq4}, i.e.,
\begin{equation} \label{eq5}
{T_{m,n}} = \mathop {{\rm{quantize}}}\limits_{{\rm{2 - bit}}} \left\{ {{\rm{angle}}\left[ {\sum\limits_{i = 1}^P {{T_{m,n,i}}} } \right]} \right\},
\end{equation}
where $\mathop {{\rm{quantize}}}\limits_{{\rm{2 - bit}}} \left\{  *  \right\}$ represents the phase bit quantization operation, which needs to be determined based on the actual full wave phase simulation results of the unit cells. ${T_{m,n}}$ is one codebook of the 3D spot beams on the RIS. Before applying the beam tracking algorithm, all codebooks will be pre-loaded into the MCU controlling the RIS board.

\subsection{RIS-based Swift Beam Tracking Scheme} 
\begin{figure}[htbp]
\centering
\includegraphics[width=3.4in]{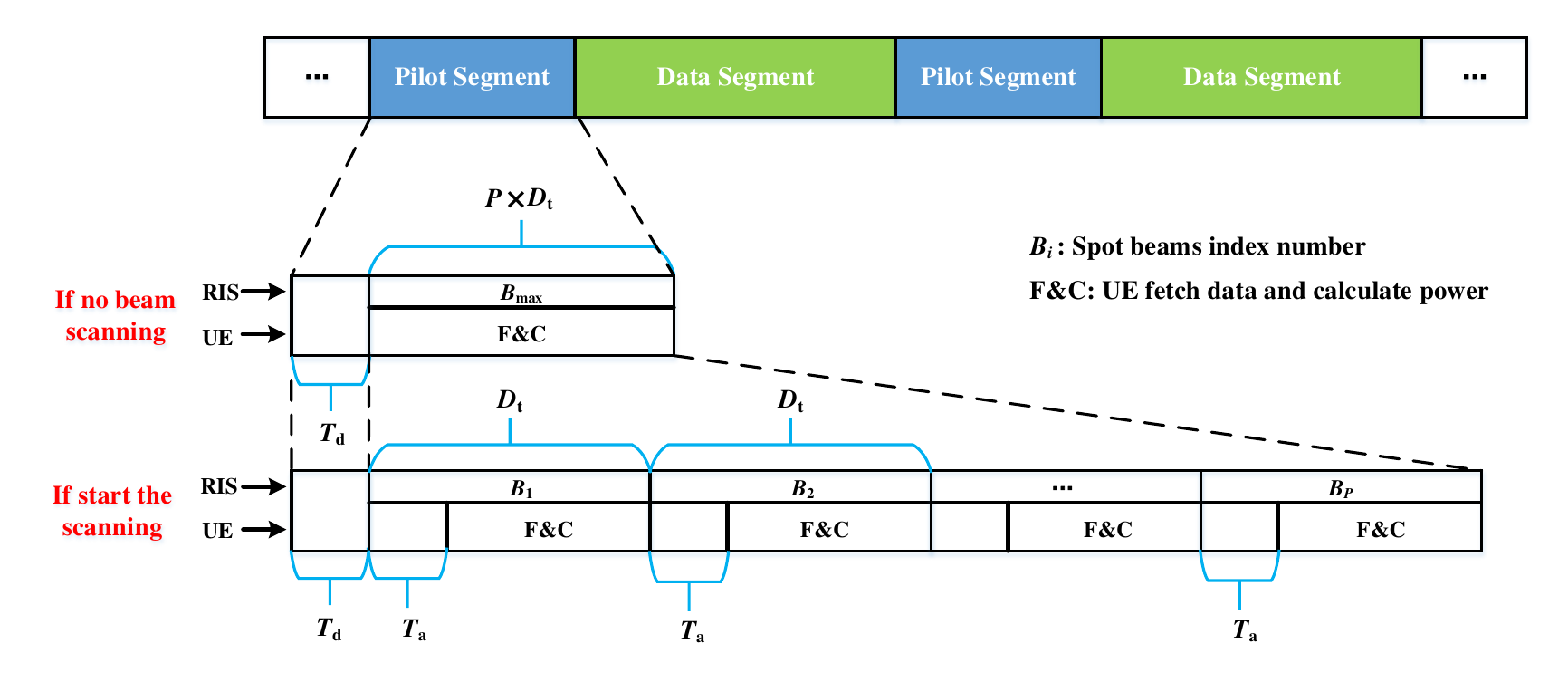}
\caption{Signal frame structure and timeline for beam scanning.}
\label{frame}
\end{figure}

In this part, we introduce a beam scanning approach aimed at tracking moving users and sustaining ongoing communication in intricate near-field settings. Spot beam scanning is a method where the RIS sequentially forms beams across each pre-defined 3D spatial region within its coverage half-space. The sequencing of the beam scanning is depicted in Fig. \ref{frame}. This process is primarily conducted during the pilot phase, where beams with varied shapes are dispatched in a time-division fashion. Every 3D spatial region is associated with a distinct scanning time slot, with the procedure repeating until the complete half-space has been covered.
% In this part, we introduce a beam scanning strategy designed to achieve beam tracking of mobile users and maintain continuous communication tasks in complex urban environments. Spot beam scanning denotes the process by which the RIS sequentially shapes beams for each pre-defined 3D spatial region within the serviced half-space it covers. The timeline for beam scanning is illustrated in Fig. \ref{frame}. The spot beam scanning process is concentrated in the pilot segment, where beams of different shapes are sent in a time-division manner. Each 3D spatial location corresponds to a specific scanning time slot, continuing until the entire half-space is traversed.

The specific process of dynamic beam tracking of the mobile user based on RIS is as follows:
\begin{itemize}
\item First, set the spatial coordinates between the transmitter and RIS, dividing the RIS's coverage into $P$ zones. The transmitter sends signals as shown in Fig. \ref{frame}. After RIS beamforming, users demodulate the signal, calculating the receiving power ${p_{\rm{r}}}$. Users set a reception power threshold ${g_{\rm{t}}}$, based on their sensitivity. If the signal power drops below this threshold, they signal the RIS to start beam scanning.
% Firstly, fix the relative spatial coordinates between the transmitter and RIS, and divide the half-space served by RIS into $P$ receiving regions. The transmitter continuously sends signals as the frame structure shown in Fig. \ref{frame}. After beamforming by RIS, the user receives and demodulates the signal in real-time, while calculating the received signal power ${p_{\rm{r}}}$. The user sets a power threshold ${g_{\rm{t}}}$ for receiving signals based on its own receiver sensitivity. Once the signal power falls below this threshold, the user feeds back information to the RIS to initiate the beam scanning process.
\item Next, for precise beam scanning, synchronization between the user and RIS on the scanning start time is crucial. Upon requesting a scan, the user also sends its local time to the RIS. The user checks its time ${T_{\rm{u}}}$ and accounts for transmission and processing delays ${t_{\rm{d}}}$, ensuring the delay $\tau$, from when the RIS starts scanning ${T_{\rm{s}}}$ to the user's time, exceeds the delay time ${T_{\rm{d}}}$. Accurate time synchronization is vital and can be achieved using the Network Time Protocol (NTP), which synchronizes devices with time sources to ms accuracy \cite{NTP}.
% Next, to ensure the accuracy of beam scanning, the user and RIS need to synchronize on the start time of the scan. When user notifies RIS to start scanning, it simultaneously sends its local time to RIS. To find the appropriate scan start time, the user first checks its local system time, denoted as ${T_{\rm{u}}}$. It's worth noting that the system's transmission and processing delays need to be tested, especially the time it takes for the user to transmit information to RIS and for RIS to respond and initiate beam scanning, denoted as ${t_{\rm{d}}}$. Here, we need to ensure that the time difference $\tau $ between the start time of RIS beam scanning ${T_{\rm{s}}}$ and the user's local time ${T_{\rm{u}}}$ is greater than ${T_{\rm{d}}}$. In this context, strict synchronization of system clocks between RIS and user is crucial, and it can be achieved through the Network Time Protocol (NTP). NTP is a widely employed protocol for synchronizing device clocks. It enables devices to synchronize with servers or time sources such as quartz clocks, GPS, etc., providing high-precision time correction with an accuracy level reaching ms level \cite{NTP}.
\item Then, at the predetermined time ${T_{\rm{s}}}$, the RIS begins to direct spot beams to the designated spatial regions, setting each beam's duration to ${D_{\rm{t}}}$. This duration, ideally, is the time needed for the RIS to adjust the reflection coefficients for its array. Starting ${T_{\rm{a}}}$ ms after ${T_{\rm{s}}}$, the user samples data, measures its power, and logs this as the reception power for that region. To ensure data capture and power calculation align with the current beam's duration, ${T_{\rm{a}}}$ must be less than ${D_{\rm{t}}}$. The user then repeats this process every ${D_{\rm{t}}}$ ms, guaranteeing that measurements for each region are taken within their respective beam durations.
% Then, at the agreed-upon time ${T_{\rm{s}}}$, the RIS starts converging spot beams onto the spatial receiving regions, with each beam's duration set to ${D_{\rm{t}}}$. Ideally, ${D_{\rm{t}}}$ represents the time required for the RIS to reconstruct the reflection coefficients for the entire array. After ${T_{\rm{a}}}$ ms from the start time ${T_{\rm{s}}}$, the user sampling partial data, calculates its power, and records it as the receiving power value for that region. To ensure that the capture and calculation operations occur within the continuous time of the current spot beam converging, ${T_{\rm{a}}}$ should be less than ${D_{\rm{t}}}$. Subsequently, the user captures data and calculates it every ${D_{\rm{t}}}$ ms to ensure that each capture occurs within the corresponding region's duration.
\item After scanning each pre-defined spatial region, the user identifies the receiving power for regions targeted by the converged spot beam. By comparing these values, the user selects the beam with the highest receiving power and informs the RIS of the selected beam ID ${B_{{\rm{ID}}}}$. The RIS then adjusts the reflection coefficients across its array to form the beam shape corresponding to ${B_{{\rm{ID}}}}$, completing the dynamic beam tracking process. The total scanning duration is influenced by the delay ${T_{\rm{d}}}$ from transmission and processing, plus the cumulative scanning time for all receiving regions, i.e.,
% Finally, after scanning all the pre-defined spatial regions, the user obtains the receiving power for each region corresponding to the converged spot beam. Through a comparative analysis, the user selects the beam corresponding to the maximum receiving power and provides feedback to the RIS with information about the beam ID ${B_{{\rm{ID}}}}$. Upon receiving this information, the RIS reconstructs the reflection coefficients for the entire array, forming the spot beam shape indicated by that beam ID ${B_{{\rm{ID}}}}$. Hence, the entire dynamic beam tracking process can be completed. The overall scanning time is jointly determined by the delay ${T_{\rm{d}}}$ introduced by transmission and processing delays and the total scanning time across all spatial regions, i.e.,
\begin{equation} \label{eq6}
    {t_{{\rm{tot}}}} = {T_{\rm{d}}} + P \times {D_{\rm{t}}},
    \end{equation}
    $P$ RIS patterns equal to $P$ spatial receiving regions.
\end{itemize}

To clarify, the steps of the RIS-based dynamic 3D beam tracking algorithm are listed as the $\textbf{Algorithm 1}$.
\begin{algorithm}[htbp]
\caption{RIS-based 3D spot beam dynamic tracking.}\label{alg:alg1}
\begin{algorithmic}[1]
\STATE {\textsc{INPUT}}: Receiving power ${p_{\rm{r}}}$, sensitivity threshold ${g_{\rm{t}}}$, user's local system time ${T_{\rm{u}}}$, duration time ${D_{\rm{t}}}$ for each spot beam, delay ${T_{\rm{a}}}$.
\STATE \hspace{0.5cm}$ \textbf{While} (1)$ \textit{\% keep monitoring the receiving power ${p_{\rm{r}}}$ and do sweeping if need}
\STATE \hspace{0.8cm}$ \textbf{If} ({p_{\rm{r}}} < {g_{\rm{t}}}) $
\STATE \hspace{1.0cm}UE Check the local system time ${T_{\rm{u}}}$;
\STATE \hspace{1.0cm}Determine the time ${T_{\rm{s}}}$ to start the beam scanning, where, ${T_{\rm{s}}} = {T_{\rm{u}}} + {T_{\rm{d}}}$;
\STATE \hspace{1.0cm}Then do the feedback, transmit ${T_{\rm{s}}}$ from the user to the RIS;
\STATE \hspace{1.0cm}Do spot beams scanning on the RIS, while calculating the receiving power on the user;
\STATE \hspace{1.0cm}$ \textbf{For} ({B_{{\rm{ID}}}} = 0; {B_{{\rm{ID}}}} = P; {B_{{\rm{ID}}}} ++) $
%\STATE \hspace{1.0cm}\% $P$ patterns equal to $P$ spatial receiving regions
\STATE \hspace{1.2cm}Delay ${T_{\rm{a}}}$ ms;
\STATE \hspace{1.2cm}Fetch signal and calculate power on the user side, and recorded as ${u_{\rm{r}}}$;
\STATE \hspace{1.2cm}Delay ${D_{\rm{t}}} - {T_{\rm{a}}}$ ms;
\STATE \hspace{1.0cm}$ \textbf{End For} $
\STATE \hspace{0.8cm}Compare and find the largest ${u_{\rm{r}}}$, recorded as ${p_{\rm{r}}}$;
\STATE \hspace{0.8cm}${B_{{\rm{max}}}} = \max {\rm{index}}\left( {{p_{\rm{r}}}} \right)$; \textit{\% Find the beam index ${B_{{\rm{max}}}}$ of the largest receiving power ar the UE side}
\STATE \hspace{0.8cm}Do the feedback of ${B_{{\rm{max}}}}$, from user to RIS;
\STATE \hspace{0.8cm}Then RIS change to the ${B_{{\rm{max}}}}$-th pattern;
\STATE \hspace{0.5cm}$ \textbf{End While} $

%\STATE \hspace{0.5cm}$ B_i \gets \sqrt{ \textsc{max}(N_{-1},N_{+1}) / N_{\mathbf{t}_i} } $ \textbf{ for } $ i = 1,...,N $
%\STATE \hspace{0.5cm}$ \hat{\mathbf{H}} \gets  B \cdot (\mathbf{X}^T\textbf{W})/( \mathbb{1}\mathbf{X} + \mathbb{1}\textbf{W} - \mathbf{X}^T\textbf{W} ) $
%\STATE \hspace{0.5cm}$ \beta \gets \left ( I/C + \hat{\mathbf{H}}^T\hat{\mathbf{H}} \right )^{-1}(\hat{\mathbf{H}}^T B\cdot \mathbf{T})  $
%\STATE \hspace{0.5cm}\textbf{return}  $\textbf{W},  \beta $
%\STATE
%\STATE {\textsc{PREDICT}}$(\mathbf{X} )$
%\STATE \hspace{0.5cm}$ \mathbf{H} \gets  (\mathbf{X}^T\textbf{W} )/( \mathbb{1}\mathbf{X}  + \mathbb{1}\textbf{W}- \mathbf{X}^T\textbf{W}  ) $
%\STATE \hspace{0.5cm}\textbf{return}  $\textsc{sign}( \mathbf{H} \beta )$
\end{algorithmic}
\label{alg1}
\end{algorithm}

This algorithm can be easily adapted to various RIS prototypes as a direct and efficient method for implementing spot beam tracking, without concern for frequency, bit quantization, or even partial components failure. The beam shapes for different 3D regions are pre-configured into the MCUs of the RIS prototyping, and the feedback mechanism is based on basic power threshold detection. This means that, in practical applications, there is no need for extensive pre-mapping or concern about random reflections affecting the codebook shapes. It can always find a codebook shape through scanning and achieve quick feedback and state locking (Although the codebook may not be optimal, it definitely ensures that the spatial beam exceeds the detection threshold at the receiver at that moment), enabling true plug-and-play functionality for the RIS system. 
The implementation cost of this mechanism involves pre-configuration of the 3D space codebooks and achieving fast switching between them. It should be noted that, in the scanning feedback process, we introduced a feedback principle based on exceeding detection thresholds rather than waiting for all codebooks to be scanned and then taking the maximum value. This significantly improves scanning efficiency and differs from traditional codebook-based schemes. Considering the inherent error correction mechanisms in communication systems, maintaining a certain detection power threshold ensures stable data transmission. This robustness is demonstrated in the following experiments, where stable tracking is achieved even with environmental disturbances.

\subsection{RIS-based Real-Time Tracking and Communication Experiments for Moving User with 3D Spot beams}
\begin{figure}[htbp]
\centering
\includegraphics[width=3.2in]{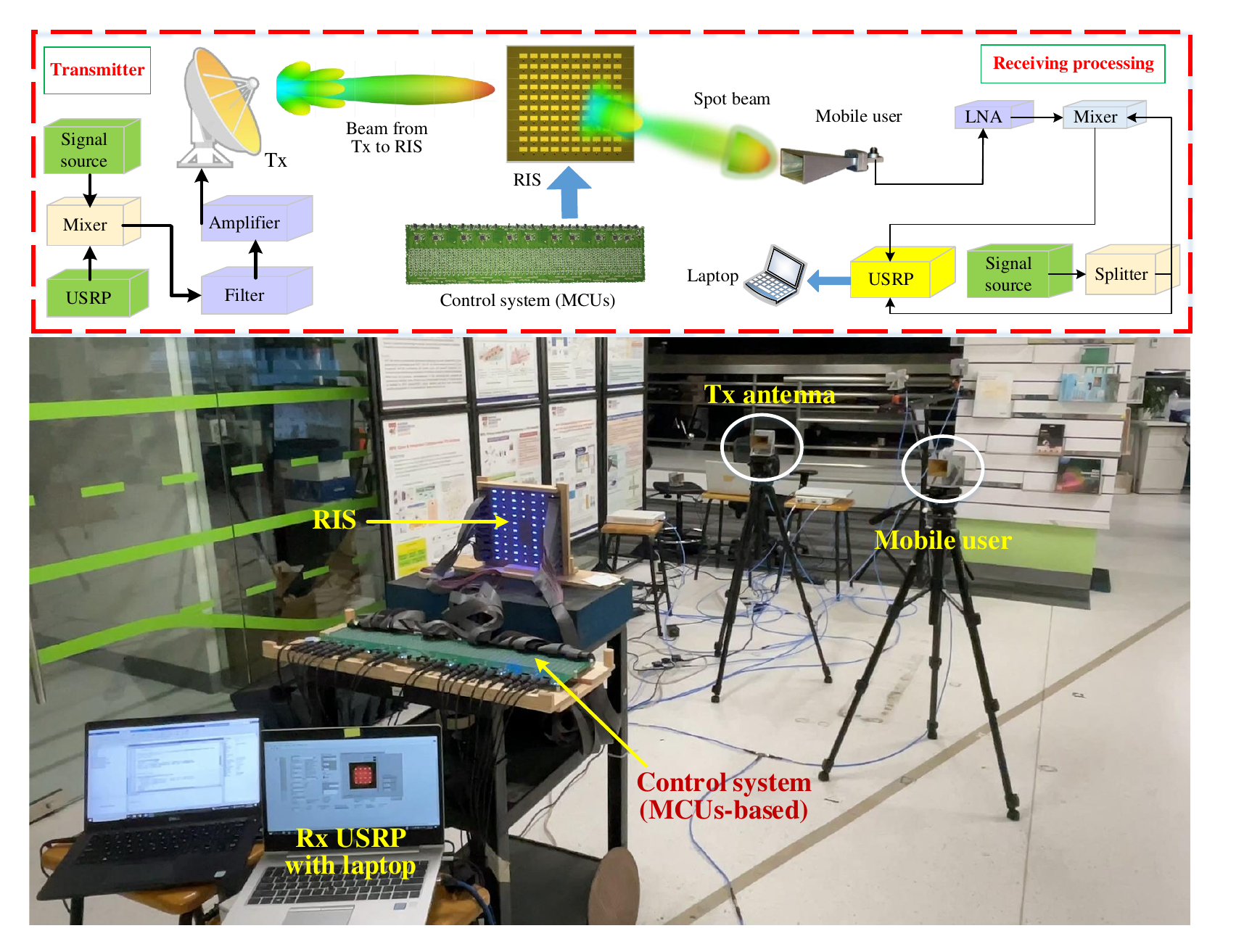}
\caption{The basic setup and on-site view of the experiment.}
\label{fig6}
\end{figure}

To verify the above algorithm, a real-time tracking and communication experiment based on RIS for the moving user has been set up in a complex environment filled with multi-path reflection interference. The experimental setup is depicted in Fig. \ref{fig6}. The base-band signal is generated by the NI USRP-2954, with a symbol rate of 125 kHz and 16-QAM modulation. The center carrier frequency of the Intermediate Frequency (IF) signal is 2.8 GHz. After frequency up-conversion by the mixer to 10.7 GHz, it is radiated into free space by a horn antenna. The transmitting antenna is positioned 2.5 meters away from the RIS. After modulation by the RIS, the transmitted signal is directionally reflected to different near-field spatial regions. Another horn antenna serves as the mobile receiving user, which will capture the RF signals reflected by the RIS. After down-conversion, another USRP-2954 is responsible for demodulation to recover the baseband data from the transmitter. The data is transmitted using single-carrier transmission without channel coding. For clarity, the main experimental parameters have been listed in Table \ref{Para}.

\begin{table}[!h]
\caption{Main experiment parameters.}
\begin{center}
\begin{tabular}{c|c|c}
\toprule%\hline
\textbf{Parameter} & \textbf{Value} & \textbf{Dimension} \\
 \midrule%\hline
 Central carrier frequency & 10.7 & GHz \\
 Intermediate frequency & 2.8 & GHz \\
 %Bits per Symbol & 4 & - \\
 Samples per symbol & 8 & - \\
 Baseband bandwidth & 125 & kHz \\
 %Roll-off factor & 1 & - \\
 Tx antenna gain & 15 & dB \\
 Transmitting power & 0 & dBm \\
 Rx antenna gain & 15 & dB \\
 Low Noise Amplifier (LNA) & 20 & dB \\
 %Airy lens diameter & 0.6 & m \\
 %Vortex topology modes & +1, -1 & - \\
 Modulation for pilot signal & 16-QAM & - \\
 Modulation for data stream & 16-QAM & - \\
 \bottomrule%\hline
\end{tabular}
\end{center}
\label{Para}
\end{table}

%one MCU works as the master and the rest work as slaves.
As illustrated in Fig. \ref{fig6}, the LabVIEW interface connected to USRPs allows the real-time observation of the demodulation constellation and Rx eye diagram. At the transmitter, a set of random sequences is cyclically transmitted as the original communication data in this experiment. The receiving user stores sufficiently long sequence data and systematically compares it with the original transmitted sequence to statistically evaluate the demodulation Bit Error Rate (BER) performance. During the experiment, the spatial position of the horn antenna is changed to simulate the random movement of the mobile user. During the movement, the RIS together with its control system dynamically changes the codebook state in real time based on the pre-defined algorithm illustrated in Sect. \ref{tracking}. Then, the shape of the reflected beams will be altered, enabling the tracking of the user's movement and achieving real-time stable data transmission (Experiment record video in https://youtu.be/riBH0YZBXtk).

It should be noted that some tracking delays were observed during the experiment, primarily attributable to the lower completion level of our prototype. In this setup, significant time was consumed by data collection and command execution processes within LabView and MATLAB. However, in the commercial product phase, the time required for the RIS to reconfigure itself would be considerably reduced, as all algorithms would already be embedded into the MCU chips. Furthermore, when the RIS is implemented in a real mobile network, it will include data packet redundancy and retransmission protocols. These mechanisms will ensure the complete reception of useful information, even in the event of brief link interruptions, thereby maintaining the actual communication experience of the mobile users.
%the transmitter sends a pre-defined random sequence using 16-QAM modulation repetitively. After up-sampling and pulse-shaping process, the baseband signal is then transmitted to 2.8 GHz by USRP and further up-converted to 10.7 GHz RF mixer and signal generator. In order to calculate the BER, the receiver only has to extract a complete sequence from the entire received data and compare it with the originally transmitted sequence. The demodulation process involves storing the received data first and then conducting off-line calculations, ensuring high stability and repeatability.
% \begin{figure}[htbp]
% \centering
% \includegraphics[width=3.2in]{interface}
% \caption{The real-time demodulation screenshot at the Rx USRP side.}
% \label{fig7}
% \end{figure}

\begin{figure}[htbp]
\centering
\includegraphics[width=3.4in]{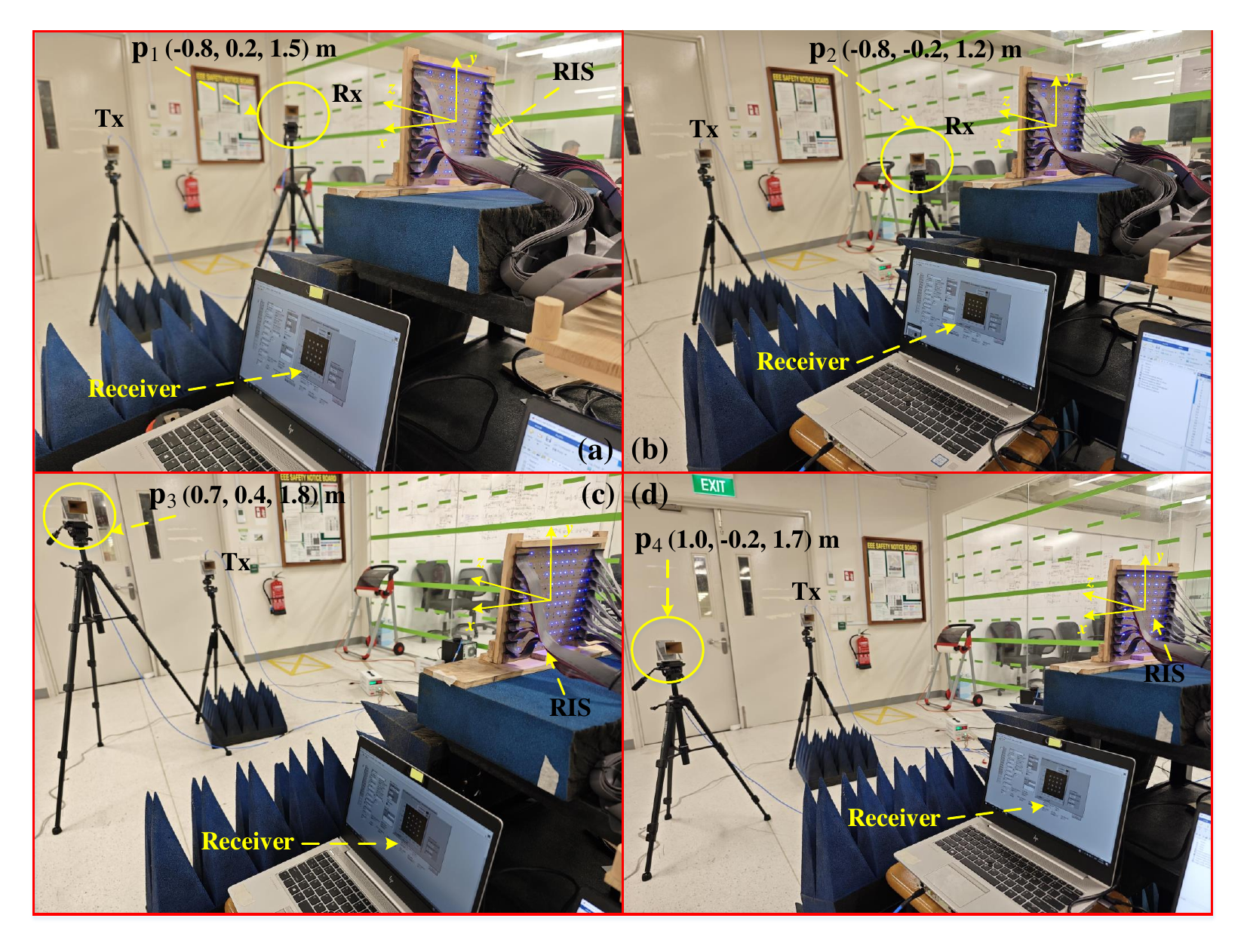}
\caption{Measurement scenarios at different 3D spatial receiving regions with respect to (a) location ${{\bf{p}}_1}$ (b) location ${{\bf{p}}_2}$ (c) location ${{\bf{p}}_3}$, and (d) location ${{\bf{p}}_4}$. (Tx antenna gain 15 dB, Tx power 0 dBm, Rx antenna gain 15 dB, Rx LNA 20 dB, bandwidth 125 kHz)}
\label{tests}
\end{figure}

\begin{figure}[htbp]
\centering
\includegraphics[width=3.4in]{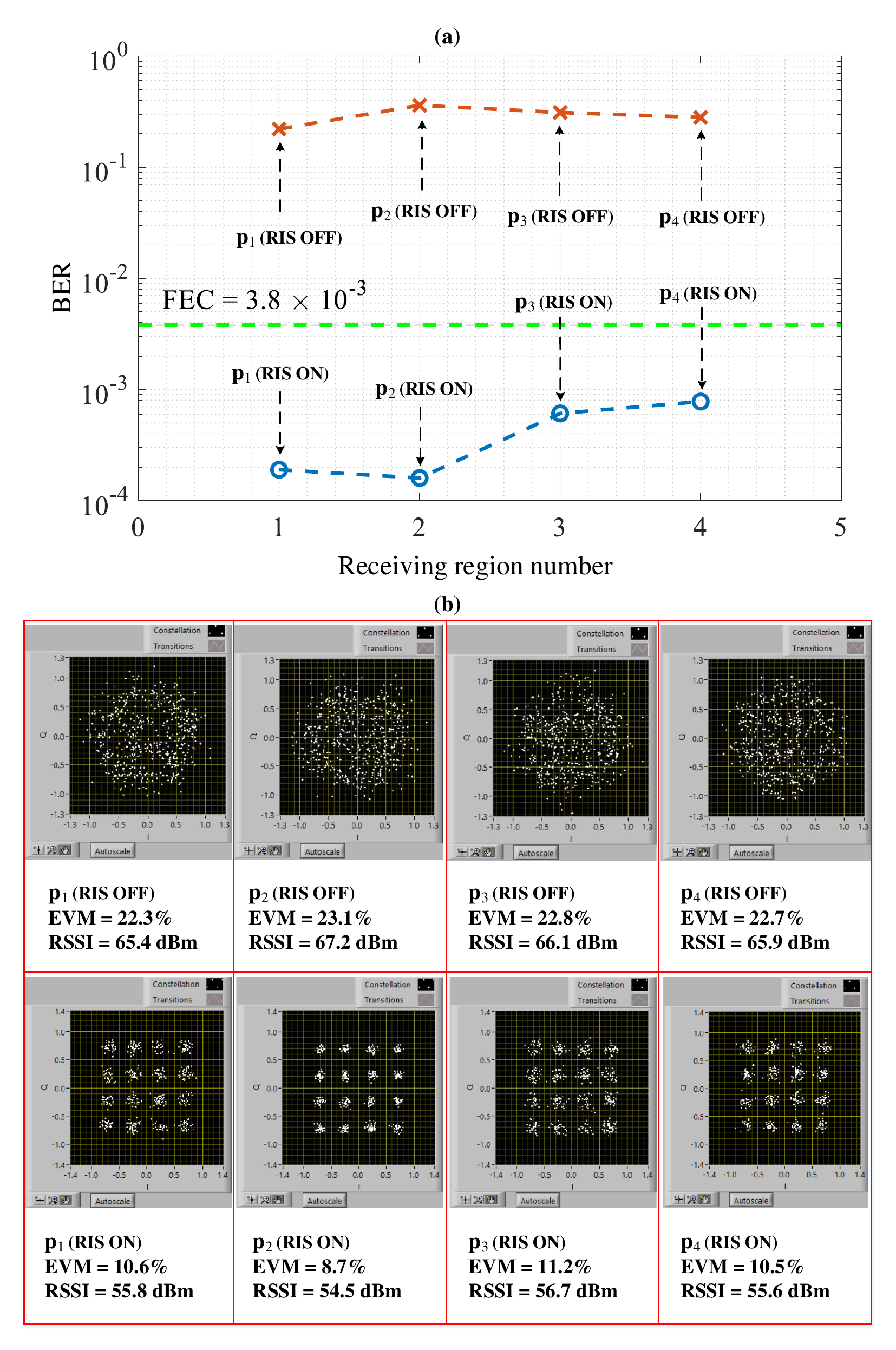}
\caption{The constellation diagrams and BER measurement results at different 3D spatial receiving regions when the RIS is OFF and ON (optimized). (a) BER comparing results at different spatial locations. (b) Constellation diagrams, EVM, and RSSI measurement results.}
\label{BER_results}
\end{figure}

Without loss of generality, considering the RIS center as the origin, we selected four spatial coordinates at varying distances and heights from the RIS to verify the 3D spot beamforming efficiency, namely, ${{\bf{p}}_1} = \left( { - 0.8,0.2,1.5} \right){\rm{m}}$, ${{\bf{p}}_2} = \left( { - 0.8,- 0.2,1.2} \right){\rm{m}}$, ${{\bf{p}}_3} = \left( {0.7,0.4,1.8} \right){\rm{m}}$, ${{\bf{p}}_4} = \left( {1.0,- 0.2,1.7} \right){\rm{m}}$. Subsequently, the RX antenna is sequentially moved to the corresponding coordinates, as shown in Fig. \ref{tests}. 
In the experiments, we measure and compare the Received Signal Strength Indicator (RSSI) under both RIS ON (optimized) and OFF states. The actual measurement results from different 3D spatial regions are recorded in Fig. \ref{BER_results}. It is obvious that when the RIS is in the OFF state or ON (optimized) state, the RSSI in the corresponding reception region differs by more than 10 dB. Then, at each of the four distinct spatial regions, the user antenna holds for a few seconds, ensuring that the receiving USRP collects and records a sufficient length of the receiving data for BER calculation. The statistical results are also depicted in Fig. \ref{BER_results}. 
Obviously, when RIS is ON (optimized) with the specific spot beam pattern, the BER statistical results at all reception regions are far below the Forward Error Correction (FEC) limit ($3.8 \times 10^{-3}$) \cite{FEC}, indicating that the RIS can automatically adjust the optimal codebook with the movement of the user, achieving beam tracking and ensuring the stability of communications. When the RIS is in the OFF state, the receiver cannot capture sufficient signal energy, resulting in a severe deterioration of the constellation diagram and BER metrics.

\section{Rotational Vortex Wavefront Manipulation through Space-Time Digital Coding on RIS Prototyping}
Moreover, to substantiate the rapid-response programmable and full-array independent control capacities of the RIS prototyping developed in this study, we utilize a phase-shifting scheme for producing time-varying rotational vortex beams. This approach leverages the unique properties of vortex beams, which possess orbital angular momentum owing to their helical wavefronts, to subtly and swiftly modulate the phase distribution throughout the RIS board. 

Vortex beams, also known as Orbital Angular Momentum (OAM) beams, are characterized by a helical phase front. The phase of the beam twists around the beam axis, resulting in a spiral wavefront. These beams carry OAM, which can be utilized for various advanced communication techniques \cite{Yufei-TAP}. The fast rotation of the wavefront in vortex beams creates a dynamic phase variation over time. When the RIS is used to generate and manipulate these vortex beams, it can induce rapid changes in the phase distribution of the reflected waves. This rapid phase variation is analogous to the Doppler shift, in other words, the rotational wavefront of the OAM beam causes the carrier phase to change more rapidly over time, resulting in a shift of the carrier's central frequency in the frequency domain. This has been verified by a lot of previous studies \cite{Malu1,Malu2,ROAM,ROAM2}. Therefore, in the Doppler domain, by rotating the vortex beam's wavefront, we can effectively modulate the signal frequency. The rotation speed of the wavefront and the OAM topological mode determine the rate of this modulation, creating a Doppler-like effect even in the absence of actual motion. This technique can be employed to modulate signals for various applications, such as enhancing communication robustness or encoding additional information in the phase of the signal.

\begin{figure*}[htbp]
\centering
\includegraphics[width=6.9in]{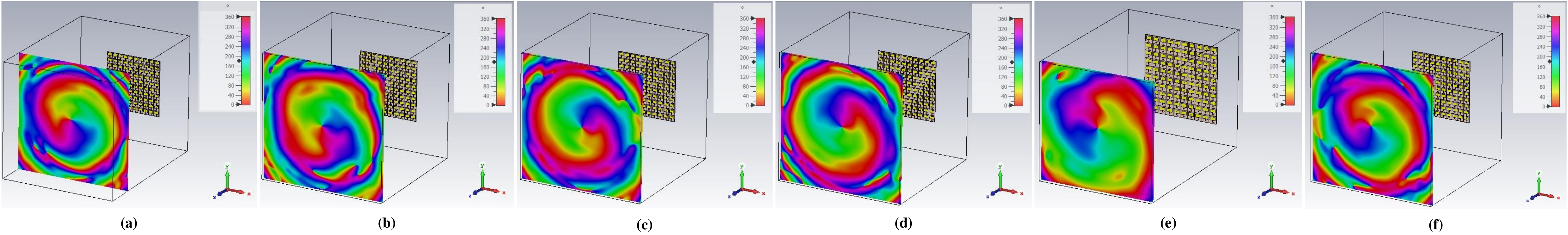}
\caption{Spinning vortex beams to generate rotational Doppler shift based highly dynamic RIS and phase-shifting scheme. (a) $l = 1$, $t = 0$. (b) $l = 1$, $t = T/6$. (c) $l = 1$, $t = 2T/6$. (d) $l = 1$, $t = 3T/6$. (e) $l = 1$, $t = 4T/6$. (f) $l = 1$, $t = 5T/6$.}
\label{Doppler}
\end{figure*}

\subsection{Rotational Vortex Beams Generation by RIS}
\begin{figure}[htbp]
\centering
\includegraphics[width=3.4in]{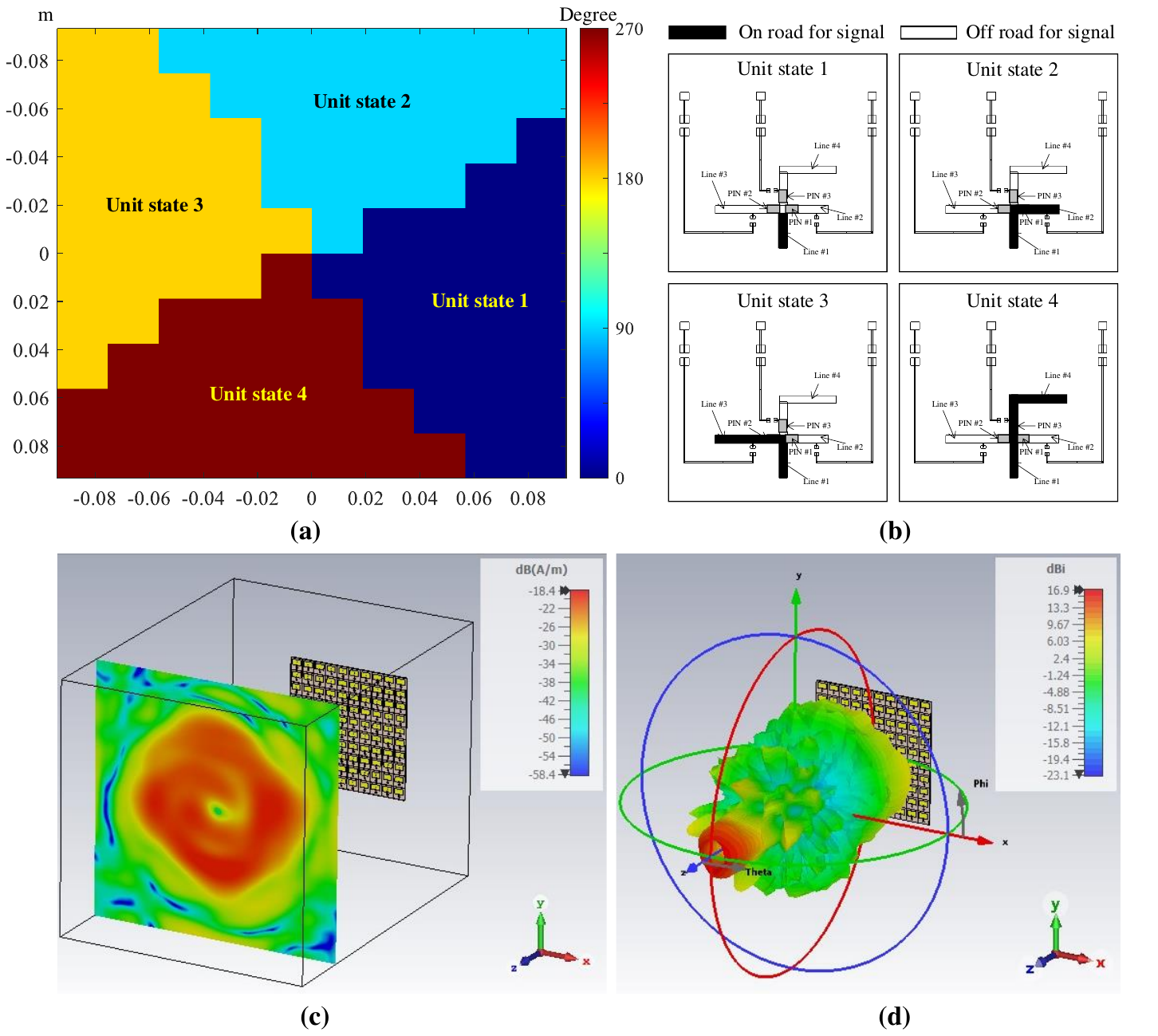}
\caption{Vortex beam generation with RIS. (a) Phase distribution on the RIS. (b) Unit states illustrations. (c) Energy distribution in a cross-section perpendicular to the beam propagation direction. (d) Radiation pattern of vortex mode 1.}
\label{RIS_vortex}
\end{figure}

As we know, by introducing an azimuthal phase factor, $\exp \left(  *  \right)$, into an EM beam, it is possible to generate an orbital angular momentum EM wave with a helical phase front. Traditional time-invariant azimuthal phase factors can be represented as $\exp \left( {jl\phi } \right)$, where $l$ denotes the topological mode of the vortex beam, and $\phi $ is the azimuthal angle. Then, the vortex beam can be expressed as,
\begin{equation} \label{eq7}
E\left( {\phi,t} \right) = {A_{\rm{e}}}\exp \left( {j2\pi {f_0}t} \right)\exp \left( {j\delta \left( \phi  \right)} \right),
\end{equation}
where ${A_{\rm{e}}}$ is the amplitude constant related to the antenna system, ${f_0}$ is the center carrier frequency. Hence, the entire metasurface can be divided into 4 distinct regions along the azimuths direction, corresponding to 4 continuously varying phase states, thereby generating vortex beams, as shown in Fig. \ref{RIS_vortex}.

By further introducing time-varying characteristics and sequentially altering the initial phase of the azimuthal phase factor across different time slots, we can achieve a spatiotemporally modulated phase, which has been illustrated in Fig. \ref{Doppler}. This modulated phase, which incorporates both spatial and temporal variations, can be expressed as,
\begin{equation} \label{eq8}
\delta \left( {\phi ,t} \right) = jl\phi  + 2\pi l \cdot t/T{\rm{ + }}\Delta,
\end{equation}
where $T$ denotes the period of rotation, $t$ is the time variable, $\Delta $ is the random initial phase. Furthermore, by differentiating \eqref{eq8} with respect to time, we can derive the expression for the Doppler frequency shift induced by the rotation of an EM beam carrying a vortex wavefront. Specifically, this can be represented as,
\begin{equation} \label{eq9}
\Delta f = \frac{1}{{2\pi }}\frac{{d\delta \left( {\phi ,t} \right)}}{{dt}} = \frac{l}{T}.
\end{equation}
It is evident that the magnitude of the Doppler frequency shift is jointly determined by the temporal modulation period $T$ of the RIS and the vortex topological mode number $l$. 

\subsection{Artificial Doppler Shift Measurement with Highly Dynamic Time-Varying Rotational Vortex Beams}
\begin{figure}[htbp]
\centering
\includegraphics[width=3.4in]{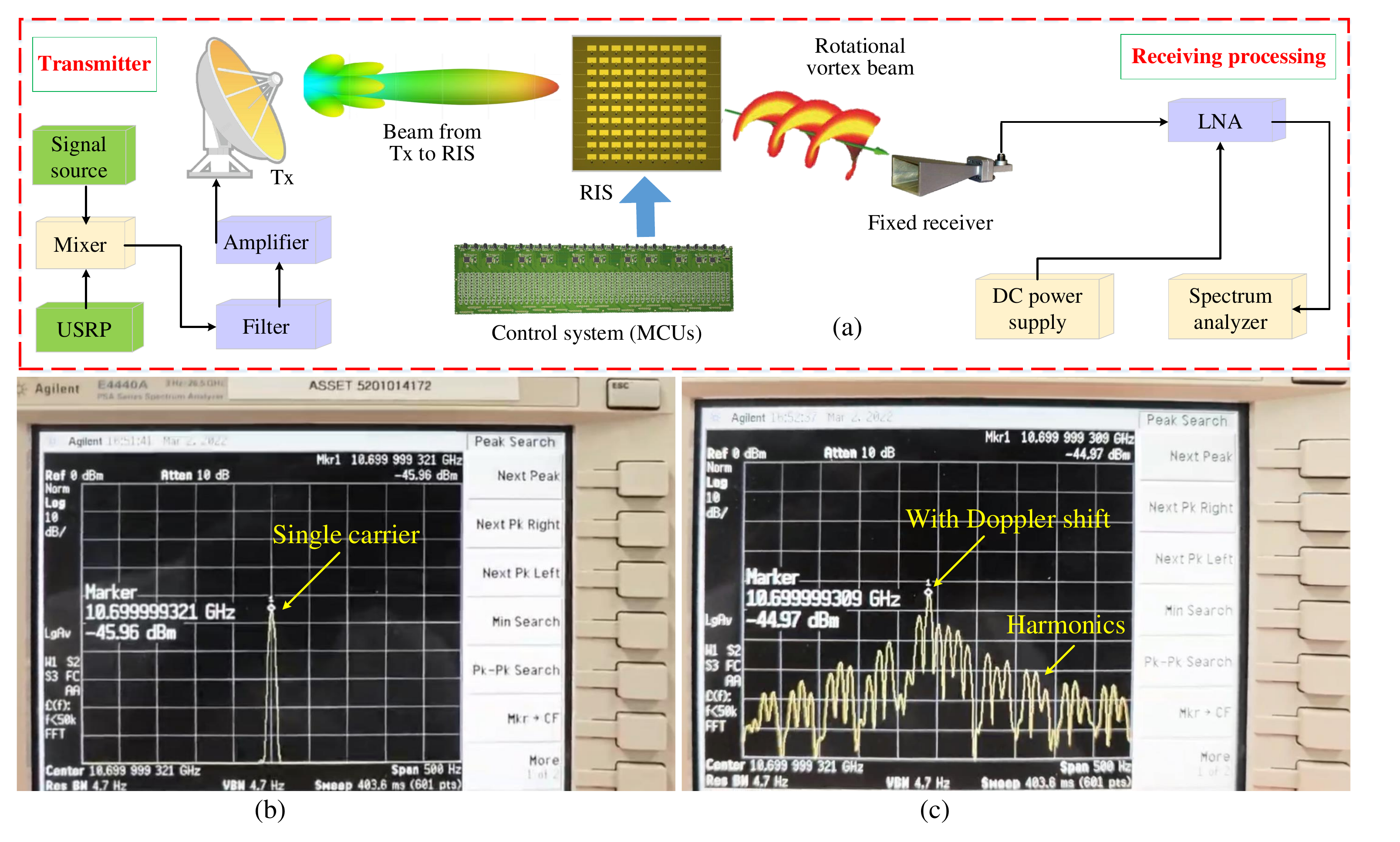}
\caption{Measured spectrum with rotational Doppler generated by the RIS prototyping. (a) Vortex beam experiment setup. (b) Single carrier without Doppler shift. (c) Single carrier with rotational Doppler shift. (Tx antenna gain 15 dB, Tx power 10 dBm, Rx antenna gain 15 dB, Rx LNA 20 dB, single carrier).}
\label{Doppler_test}
\end{figure}

Through precise alterations of the time-varying vortex beams generated by the RIS, we achieve dynamic control over artificial Doppler shifts within the signal frequency domain. This capability not only demonstrates the RIS prototyping's highly dynamic beamforming potential but also highlights its utility in enhancing various application scenarios through sophisticated manipulation of EM waves, e.g., Doppler radar velocity deception and stealth \cite{ROAM,ROAM2}. A specific example has been illustrated in Fig. \ref{Doppler}, which describes how we rotated the vortex beams in detail, and the artificial Doppler experiment setup and measurement results are shown in Fig. \ref{Doppler_test} (Video recorded in https://youtu.be/IJDO-I-umaI).

% In our experiments, we used an Agilent E4440A spectrum analyzer to measure the center frequency of a stationary single-carrier signal and the frequency spectrum when the RIS rapidly changes states, as shown in Fig. \ref{Doppler_test}(a)(b). We observed a center frequency shift of $\Delta f = 12$ Hz during rapid vortex wavefront rotation by the RIS, indicating millisecond-level dynamic beam control. Switching among 12 vortex states ($l = 1$) suggests a switching time of approximately $\frac{1}{{12 \times \Delta f}}$ seconds, aligning with Section II B's calculations. This experiment confirms our RIS's precision in controlling EM beams and shows how adjusting modulation parameters and vortex modes can finely control the Doppler shift for sophisticated wireless tasks.
In our experiments, we utilized a spectrum analyzer (Agilent E4440A) to measure the center frequency of the single-carrier signal reflected by the RIS in a stationary state, as well as the frequency spectrum of the signal when the RIS was in space-time digital-coding states. The comparative results of their spectra are illustrated in Fig. \ref{Doppler_test}(a)(b). It is distinctly observable that when the RIS generates a vortex undergoing rapid wavefront rotation through space-time digital-coding, the center frequency of the carrier signal shifts by $\Delta f = 12$ Hz. In the experiment, we employed rapid switching among 12 different vortex wavefront states ($l = 1$) continuously, implying that the time required for the RIS to switch between each vortex state is approximately $\frac{1}{{12 \times \Delta f}}$ seconds. This observation aligns with the calculations presented in Section II B, further corroborating the rapid-response control capabilities of our RIS prototype for EM beam manipulation. Moreover, this experiment demonstrates that by adjusting the temporal modulation parameters and selecting specific vortex modes, it is possible to finely tune the Doppler shift.
% , offering a nuanced approach to wavefront modulation for advanced wireless applications.
This experimental evidence highlights the prototyping's advanced functionality in precisely manipulating electromagnetic waves, demonstrating its potential for high-speed, dynamic wireless communication applications.
This research blurs the boundary between RF antennas and digital signal processing, offering a new perspective for studying encoded EM environment signal processing. Some new digital signal processing algorithms can be designed based on this finding, e.g., OTFS technology applied in low mobility scenarios based on artificial Doppler shift generated by RIS-OAM system \cite{APMC-ge}.

\section{Outdoor Far-Field Beamforming Testing under Blocking Scenarios}
\begin{figure}[htbp]
\centering
\includegraphics[width=3.3in]{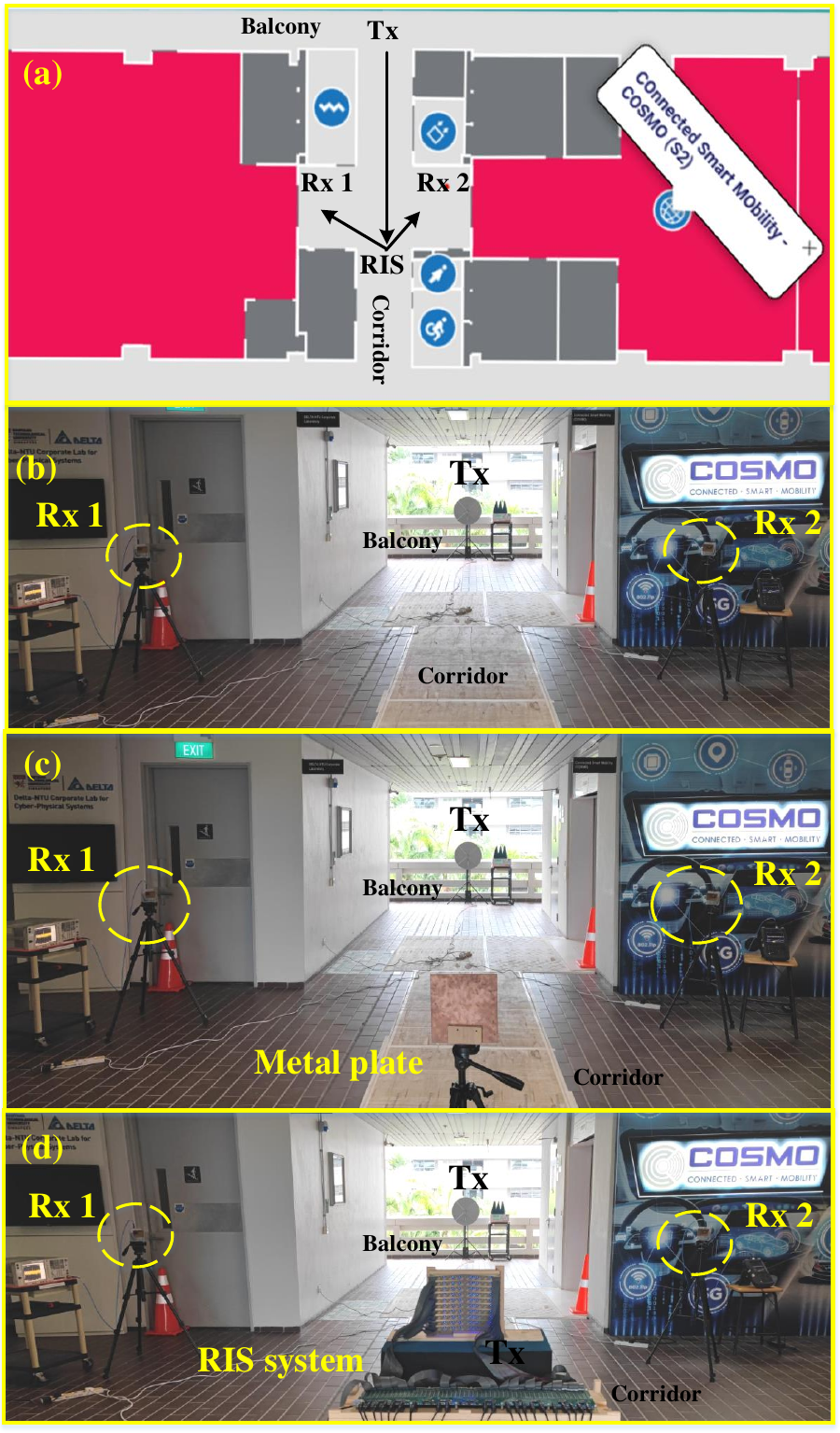}
\caption{RIS comparative experiments in the outdoor far-field scenario. (a) Diagram of the experimental setup. (b) The line-of-sight corridor without reflective objects. (c) The line-of-sight corridor featuring a metal plate of the same size as the RIS to enhance signal reflection. (d) The line-of-sight corridor utilizing the RIS to modulate signal reflection. (Tx reflector antenna gain 30 dB, Tx power 20 dBm, Rx antenna gain 15 dB, Rx LNA 20 dB, single carrier)}
\label{scenario2}
\end{figure}

Furthermore, for outdoor far-field application scenarios, our RIS prototyping also demonstrates its ability to efficiently distribute energy to specified areas through the manipulation of beam directions. This capability is particularly important for non-line-of-sight users obstructed by buildings. To validate the capability of the RIS board with this $10 \times 10$ size to modulate radio waves in natural environments, we conducted a comparative validation experiment on an outdoor balcony and corridor near our lab, as shown in Fig. \ref{scenario2}. This approach ensures that the RIS's effectiveness in manipulating EM waves is tested under conditions that closely mimic real-world scenarios, thus providing a robust assessment of its practical application potential in outdoor wireless scenarios. To mitigate interference with other commercial frequency bands, all the outdoor experiments are operated in the X-band. The transmission setup features a reflector antenna that emits a single-carrier signal directly from the RF signal source with an extremely narrow beam illuminating the RIS board. This signal follows a line-of-sight path to the RIS, which then reflects it toward receivers located in the corners of the corridor. At the receiving end, a standard-gain directional horn antenna captures the signal. Notably, due to obstructions like wall corners and the directional characteristics of the antennas, there is no direct line-of-sight connection between the transmitter and receiver.
% The transmitting end utilizes a reflector antenna to generate a single-carrier signal directly from the RF signal source. This signal traverses a line-of-sight path to the RIS, where it is subsequently reflected towards the receivers situated in corners of the corridor. The receiving end employs a standard-gain directional horn antenna to capture the signal. Notably, due to obstructions such as wall corners and the directional nature of the antennas, there is no direct line-of-sight link between the receiver and the transmitter.

To provide a comparative analysis with the RIS-aided communication link, we set up two additional scenarios as control groups, as depicted in Fig. \ref{scenario2}(b) and \ref{scenario2}(c). In scenario Fig. \ref{scenario2}(b), no reflective objects are placed directly in front of the transmitter. The signal is emitted by the Tx antenna from a balcony on one side of the building, traverses the corridor, and radiates into free space from a balcony on the opposite side. In contrast, scenario Fig. \ref{scenario2}(c) features a metal plate of similar dimensions positioned at the same location as the RIS along the direct signal propagation path. Upon reflecting off the metal plate, the energy of the signal is captured by the receiving antenna at both mirrored and non-mirrored reflection directions, which is then compared to the scenario with an RIS present, as shown in Fig. \ref{scenario2}(d). At the receiver's location, a spectrum analyzer is employed to record and compare the variations in reflected signal energy. The experimental results are tabulated in Table \ref{tab1}.
\begin{table}[htbp]
\caption{Receiving power captured by the microwave analyzer (Keysight, N9917A) in testing scenarios shown by Fig. \ref{scenario2}.}
\begin{center}
\begin{tabular}{c|c|c}
\toprule%\hline
\textbf{Scenarios} & \textbf{Rx 1} & \textbf{Rx 2} \\
 \midrule%\hline
 No reflective objects & -61.8 dBm & -60.7 dBm \\
 Rx 1 at about $0^\circ $ to metal plate & -45.1 dBm & NA \\
 Rx 1 move $15^\circ $ to metal plate & -54.7 dBm & NA \\
 RIS beam sweep to Rx 1 ($+45^\circ, 4 m$) & -46.6 dBm & -55.2 dBm \\
 RIS beam sweep to Rx 2 ($-30^\circ, 4 m$) & -54.3 dBm & -45.8 dBm \\
 \bottomrule%\hline
\end{tabular}
\end{center}
\label{tab1}
\end{table}
% \begin{table}[!t]
% 	\renewcommand{\arraystretch}{1.3}
% 	\caption{Receiving power captured by the microwave analyzer (Keysight, N9917A).}
% 	\centering
% 	\label{tab1}
% 	%\centering
% 	\resizebox{\columnwidth}{!}{
% 		\begin{tabular}{l l l}
% 			\hline\hline \\[-3mm]
% 			\multicolumn{1}{c}{Scenarios} & \multicolumn{1}{c}{Rx 1} & \multicolumn{1}{c}{Rx 2}  \\[1.6ex] \hline
% 			No reflective objects & -61.8 dBm & -60.7 dBm \\
% 			Rx 1 at about $0^\circ $ to metal plate & -45.1 dBm & NA \\ 
% 			Rx 1 move $15^\circ $ to metal plate & -54.7 dBm & NA \\ 
% 			RIS beam sweep to Rx 1 ($+45^\circ, 4$ m) & -46.6 dBm & -55.2 dBm \\
% 			IRS beam sweep to Rx 2 ($-30^\circ, 4$ m) & -54.3 dBm & -45.8 dBm \\ [1.4ex]
% 			\hline\hline
% 		\end{tabular}
% 	}
% \end{table}

From experimental results, it's evident that in the absence of any reflective objects along the direct line-of-sight path, the receiver can only receive the signal power around -60 dBm by random reflections. When a metal plate is introduced into this direct path, a noticeable increase in received power to approximately -45.1 dBm is observed along the normal direction of the metal plate. However, as one deviates from this normal direction, the received signal power rapidly diminishes. This suggests that the metal plate can only provide specular reflection and is not adaptable to mobile communication environments. Subsequently, we replaced the metal plate with an RIS panel of equivalent dimensions and positioned the receiver at angles of $-30^\circ, 4$ m and $45^\circ, 4$ m relative to the RIS. By employing MCUs-controlled RIS to steer reflected beams to various locations, significant variations in received power were observed, as detailed in Table \ref{tab1}.
%\begin{figure}[htbp]
%\centering
%\includegraphics[width=3.0in]{Constellation}
%\caption{Demodulation constellation diagrams. (a) Rx 1 with the directional focus of the IRS beam. (b) Rx 1 without the directional focus of the IRS beam, signal only getting from random reflection of the environment.}
%\label{fig6}
%\end{figure}

For instance, when the RIS adjusts the direction of the reflected beam towards $+45^\circ $, the received power at Rx 1 is approximately -46.6 dBm. Conversely, when the beam is steered towards a non-mirror-symmetric angle, i.e. $-30^\circ $ direction, the received power at Rx 1 direction drops by about 10 dB, while the received power at Rx 2 direction elevates to -45.8 dBm. Similarly, when the RIS-controlled beam is redirected from Rx 2 towards Rx 1, which is not mirror-symmetric, the received power at Rx 1 increases, while the received power at Rx 2 decreases to 55.2 dBm. It is evident that higher receiving power leads to an improved signal-to-noise ratio, thereby enhancing the demodulation performance at the receiver. This not only demonstrates that our RIS prototyping offers flexible beamforming capabilities, but also proves that even under outdoor far-field conditions, an RIS board of this size can still provide a considerable increase in received power for obstructed users. This has significant implications for deep coverage and ensuring communication quality in mobile communication scenarios.

It is worth emphasizing that while a simple metal reflector can provide basic signal reflection, RIS technology offers significant advantages that justify its practicality and necessity. RIS provides dynamic beamforming and steering, allowing it to adjust the phase and amplitude of reflected signals in real-time, optimize signal strength, and coverage dynamically. Additionally, RIS actively mitigates interference, improves energy efficiency through targeted reflection, and supports advanced functionalities like secure communication, environmental sensing, and spatial modulation. Our experiments in Fig. \ref{scenario2} and Table \ref{tab1} demonstrate that even a small RIS prototype can enhance signals effectively in practical scenarios, proving its superiority in dynamic and varied real-world applications.
% This demonstrates that our RIS prototyping offers flexible beamforming capabilities, serving as a crucial bridge between signal processing algorithms and the actual EM environment.

\section{Conclusion and Discussion}
This paper has unveiled an exhaustive exploration into the design, implementation, and experimental validation of an innovative 2-bit reconfigurable metasurface prototyping, characterized by its capability for full-array flexible control and rapid-response to control inputs. This prototyping facilitates a plethora of wireless solutions, including 3D spot beam tracing, artificial rotational Doppler modulation by time-varying vortex beams, and outdoor far-field beam steering. Notably, the prototyping achieves rapid switching of the entire RIS array's patterns of less than 1 millisecond, which may enable it to track fast-moving targets such as small UAVs, bicycles, and cars navigating urban roads in the future, and offer a new perspective for studying encoded EM environment signal processing.

Central to our investigation is the introduction of a novel 2-bit phase-quantized unit cell design which is distinguished by its capability to maintain 4 stable phase states with ${90^ \circ }$ differences, an energy attenuation of less than 0.6 dB across a bandwidth of 200 MHz, and full-array flexible independent control. By achieving such low insertion loss while providing multiple phase configurations, our prototyping offers a robust platform for the development of advanced wireless solutions. Supported by a programmable hardware architecture and enhanced by the parallel processing and pre-configuration capabilities of meticulously designed multi-MCU control circuits, this system embodies the epitome of rapid and precise phase adjustment capabilities. This foundation enables the execution of complex structured EM waves, allowing for the dynamic manipulation of the EM wavefront to meet the demanding requirements of modern wireless technologies. Moreover, this study develops a tailored spot beamforming codebook scheme for RIS-aided 3D spot beam tracking. Introducing an adaptable and efficient RIS scanning algorithm, this system facilitates the dynamic and robust communication link for moving users.
% , significantly enhancing the adaptability of 3D spot beam tracking. 
% Introducing an adaptable and efficient spot beam scanning algorithm, this system facilitates the rapid tracking of moving users by comprehensively scanning predetermined spatial regions.

Reflecting on this research, we recognize that the contributions of this paper extend the horizons of reconfigurable metasurface technologies, offering fresh perspectives for future explorations, e.g., integration of RIS with existing wireless standards, high-efficient wireless power transfer, integrated sensing and communication, etc. The promise held within the pages of this paper is but a glimpse into the future, where the confluence of RIS and wireless technologies transforms the landscape of how we connect, power, and perceive the world around us.


\begin{thebibliography}{00}
% \bibitem{b50} S. Azodolmolky, Jordi Perell\'{o}, Marianna Angelou, Fernando Agraz, Luis Velasco, Salvatore Spadaro,~\textit{et al.}, Experimental demonstration of an impairment aware network planning and operation tool for transparent/translucent optical networks,''~\textit{J. Lightwave. Technol.}, vol. 29, no. 4, pp. 439--448, Sep. 2011.

\bibitem{open1} 
C. Huang, S. Hu, G. C. Alexandropoulos, A. Zappone, C. Yuen, R, Zhang, M. D. Renzo, and M. Debbah, ``Holographic MIMO surfaces for 6G wireless networks: Opportunities, challenges, and trends,''~\textit{IEEE Wireless Communications}, vol. 27, no. 5, pp. 118-125, Oct. 2020, doi: 10.1109/MWC.001.1900534.

\bibitem{open2} 
P. Mursia, F. Devoti, V. Sciancalepore, and X. C. -P$\Acute{e}$rez, ``RISe of Flight: RIS-empowered UAV communications for robust and reliable air-to-ground networks,''~\textit{IEEE Open Journal of the Communications Society}, vol. 2, pp. 1616-1629, 2021, doi: 10.1109/OJCOMS.2021.3092604.

\bibitem{open3} 
J. Wang, W. Tang, Y. Han, S. Jin, X. Li, C. -K. Wen, Q. Cheng, and T. J. Cui, ``Interplay between RIS and AI in wireless communications: Fundamentals, architectures, applications, and open research problems,''~\textit{IEEE Journal on Selected Areas in Communications}, vol. 39, no. 8, pp. 2271-2288, Aug. 2021, doi: 10.1109/JSAC.2021.3087259.

% \bibitem{qingqing} 
% Q. Wu and R. Zhang, ``Towards smart and reconfigurable environment: Intelligent reflecting surface aided wireless network,''~\textit{IEEE Communications Magazine}, vol. 58, no. 1, pp. 106-112, Jan. 2020, doi: 10.1109/MCOM.001.1900107.

% \bibitem{open4}
% K. Shafique and M. Alhassoun, ``Going beyond a simple RIS: trends and techniques paving the path of future RIS,''~\textit{IEEE Open Journal of Antennas and Propagation}, vol. 5, no. 2, pp. 256-276, Apr. 2024, doi: 10.1109/OJAP.2024.3360900.

\bibitem{open5}
S. Kisseleff, W. A. Martins, H. Al-Hraishawi, S. Chatzinotas, and B. Ottersten, ``Reconfigurable intelligent surfaces for smart cities: Research challenges and opportunities,''~\textit{IEEE Open Journal of the Communications Society}, vol. 1, pp. 1781-1797, 2020, doi: 10.1109/OJCOMS.2020.3036839.

\bibitem{cui1}
Z. X. Wang, H. Q. Yang, F. Zhai, J. W. Wu, Q. Cheng, and T. J. Cui, ``A low-cost and low-profile electronically programmable bit array antenna for two-dimensional wide-angle beam steering,''~\textit{IEEE Transactions on Antennas and Propagation}, vol. 71, no. 1, pp. 342-352, Jan. 2023, doi: 10.1109/TAP.2022.3221840.

\bibitem{long1}
L. Zhu, J. Han, G. Li, X, Ma, D. Xia, Z. Zheng, H. Liu, and L. Li, ``Dual Linearly Polarized 2-bit Programmable Metasurface With High Cross-Polarization Discrimination,''~\textit{IEEE Transactions on Antennas and Propagation}, vol. 72, no. 2, pp. 1510-1520, Feb. 2024, doi: 10.1109/TAP.2023.3345034.

\bibitem{dai1}
J. Zhu, K. Liu, Z. Wan, L. Dai, T. J. Cui, and H. V. Poor, ``Sensing RISs: Enabling dimension-independent CSI acquisition for beamforming,''~\textit{IEEE Transactions on Information Theory}, vol. 69, no. 6, pp. 3795-3813, Jun. 2023, doi: 10.1109/TIT.2023.3243836.

\bibitem{R2-2}
S. Yue, S. Zeng, H. Zhang, F. Lin, L. Liu, and B. Di, ``Intelligent omni-surfaces aided wireless communications: Does the reciprocity hold?''~\textit{IEEE Transactions on Vehicular Technology}, vol. 72, no. 6, pp. 8181-8185, Jun. 2023, doi: 10.1109/TVT.2023.3242283.

\bibitem{Access}
H. Zhang, X. Chen, Z. Wang, Y. Ge, and J. Pu, ``A 1-bit electronically reconfigurable reflectarray antenna in X band,''~\textit{IEEE Access}, vol. 7, pp. 66567-66575, May 2019, doi: 10.1109/ACCESS.2019.2918231.

\bibitem{Yang}
M. Wang, S. Xu, F. Yang, and M. Li, ``A 1-bit bidirectional reconfigurable transmit-reflect-array using a single-layer slot element with PIN diodes,''~\textit{IEEE Transactions on Antennas and Propagation}, vol. 67, no. 9, pp. 6205-6210, Jul. 2019, doi: 10.1109/TAP.2019.2925925.

\bibitem{bit1}
H. Luyen, J. H. Booske, and N. Behdad, ``2-Bit phase quantization using mixed polarization-rotation/non-polarization-rotation reflection modes for beam-steerable reflectarrays,''~\textit{IEEE Transactions on Antennas and Propagation}, vol. 68, no. 12, pp. 7937-7946, Dec. 2020, doi: 10.1109/TAP.2020.3000517.

\bibitem{bit2}
X. Hu, R. Deng, B. Di, H. Zhang, L. Song, ``Holographic beamforming for ultra massive MIMO with limited radiation amplitudes: How many quantized bits do we need?,''~\textit{IEEE Communications Letters}, vol. 26, no. 6, pp. 1403-1407, Jun. 2022, doi: 10.1109/LCOMM.2022.3151801.

\bibitem{chengqiang}
L. Zhang, X. Q. Chen, R. W. Shao, J. Y. Dai, Q. Cheng, G. Castaldi, V. Galdi, and T. J. Cui, ``Breaking Reciprocity with Space-Time-Coding Digital Metasurfaces,''~\textit{Advanced Materials}, vol. 31, no. 41, p. 1904069, Oct. 2019, doi: 10.1002/adma.201904069.

\bibitem{control}
H. Kamoda, T. Iwasaki, J. Tsumochi, T. Kuki, and O. Hashimoto, ``60-GHz electronically reconfigurable large reflectarray using single-bit phase shifters,''~\textit{IEEE Transactions on Antennas and Propagation}, vol. 59, no. 7, pp. 2524-2531, Jul. 2011, doi: 10.1109/TAP.2011.2152338.

\bibitem{nano}
L. Wang, Y. Zhang, X. Guo, ~\textit{et al.}, ``A review of THz modulators with dynamic tunable metasurfaces,''~\textit{Nanomaterials}, vol. 9, no. 7, p. 965, Jul. 2019, doi: 10.3390/nano9070965.

\bibitem{resonance}
J. Liao, S. Guo, L. Yuan, C. Ji, C. Huang, and X. Luo, ``Independent manipulation of reflection amplitude and phase by a single-layer reconfigurable metasurface,''~\textit{Advanced Optical Materials}, vol. 10, no. 4, p. 2101551, Dec. 2021, doi: 10.1002/adom.202101551.

\bibitem{proceeding}
Q. Cheng, L. Zhang, J. Y. Dai,~\textit{et al.}, ``Reconfigurable intelligent surfaces: Simplified-architecture transmitters - From theory to implementations,''~\textit{Proceedings of the IEEE}, vol. 110, no. 9, pp. 1266-1289, Sept. 2022, doi: 10.1109/JPROC.2022.3170498.

\bibitem{arxiv}
C. Liu, F. Yang, S. Xu, M. Li, ``Reconfigurable metasurface: A systematic categorization and recent advances,''~\textit{arXiv}, Jan. 2023, doi: 10.48550/arXiv.2301.00593.

\bibitem{theory}
C. -C. Cheng and A. A. -Tamijani, ``Study of 2-bit antenna-filter-antenna elements for reconfigurable millimeter-wave lens arrays,''~\textit{IEEE Transactions on Microwave Theory and Techniques}, vol. 54, no. 12, pp. 4498-4506, Dec. 2006, doi: 10.1109/TMTT.2006.885993.

\bibitem{Deng}
Y. Yin, C. Deng, X Cao, Y. Hao, Y. Hou, H Xu, and K. Sarabandi, ``Design of a 2-bit dual-polarized reconfigurable reflectarray with high aperture efficiency,''~\textit{IEEE Transactions on Antennas and Propagation}, vol. 72, no. 1, pp. 542-552, Jan. 2024, doi: 10.1109/TAP.2023.3326951.

\bibitem{EuCAP}
L. G. da Silva, P. Xiao, and A. C. S., ``A 2-bit tunable unit cell for 6G reconfigurable intelligent surface application,''~\textit{2022 16th European Conference on Antennas and Propagation (EuCAP)}, Madrid, Spain, 2022, pp. 1-5, doi: 10.23919/EuCAP53622.2022.9769482.

\bibitem{R1-2}
S. Zeng, B. Di, H. Zhang, J. Gao, S. Yue, X. Hu, R. Fu, J. Zhou, X. Liu, H. Zhang, Y. Wang, S. Sun, H. Qin, X. Su, M. Wang, and L. Song, ``RIS-based IMT-2030 testbed for mmwave multi-stream ultra-massive MIMO communications,''~\textit{IEEE Wireless Communications}, Early Access Article, 2024, doi: 10.1109/MWC.005.2300052.
% B. Di, H. Zhang, J. Gao, S. Yue, X. Hu, R. Fu, J. Zhou, X. Liu, H. Zhang, Y. Wang, S. Sun, H. Qin, X. Su, M. Wang, and L. Song

\bibitem{R2-4}
L. Dai, B. Wang, M. Wang, X. Yang, J. Tan, S. Bi, S. Xu, F. Yang, Z. Chen, M. D. Renzo, C.-B. Chae, and L. Hanzo, ``Reconfigurable intelligent surface-based wireless communications: Antenna design, prototyping, and experimental results,'' ~\textit{IEEE Access}, vol. 8, pp. 45913-45923, 2020, doi: 10.1109/ACCESS.2020.2977772.

\bibitem{R1-1}
M. Rossanese, P. Mursia, A. G.-Saavedra, V. Sciancalepore, A. Asadi, and X. C.-Perez, ``Designing, building, and characterizing RF switch-based reconfigurable intelligent surfaces,'' In Proceedings of the 16th ACM Workshop on Wireless Network Testbeds, Experimental Evaluation \& Characterization (WiNTECH'22). Association for Computing Machinery, New York, NY, USA, 2022, pp. 69-76, doi: 10.1145/3556564.3558236.
% B. Wang, M. Wang, X. Yang, J. Tan, S. Bi, S. Xu, F. Yang, Z. Chen, M. D. Renzo, C.-B. Chae, and L. Hanzo

\bibitem{R3-4}
J. -B. Gros, V. Popov, M. A. Odit, V. Lenets, and G. Lerosey, ``A reconfigurable intelligent surface at mmwave based on a binary phase tunable metasurface,''~\textit{IEEE Open Journal of the Communications Society}, vol. 2, pp. 1055-1064, 2021, doi: 10.1109/OJCOMS.2021.3076271.

\bibitem{R4-4}
R. Wang, Y. Yang, B. Makki, and A. Shamim, ``A wideband reconfigurable intelligent surface for 5G millimeter-wave applications,''~\textit{IEEE Transactions on Antennas and Propagation}, vol. 72, no. 3, pp. 2399-2410, March 2024, doi: 10.1109/TAP.2024.3352828.

\bibitem{R5-4}
J. Zhao,~\textit{et al.}, ``Programmable time-domain digital-coding metasurface for non-linear harmonic manipulation and new wireless communication systems,''~\textit{Nat. Sci. Rev.}, vol. 6, no. 2, pp. 231-238, Mar. 2019, doi: 10.1093/nsr/nwy135.

\bibitem{R7-4}
J. Y. Dai, J. Zhao, Q. Cheng, and T. J. Cui, ``Independent control of harmonic amplitudes and phases via a time-domain digital coding metasurface,''~\textit{Light Sci. Appl.}, vol. 7, no. 1, p. 90, 2018, doi: 10.1038/s41377-018-0092-z.


\bibitem{APMC}
L. Zhang and T. J. Cui, ``Angle-insensitive 2-bit programmable coding metasurface with wide incident angles,''~\textit{2019 IEEE Asia-Pacific Microwave Conference (APMC)}, Singapore, 2019, pp. 932-934, doi: 10.1109/APMC46564.2019.9038764.

\bibitem{Dai2}
M. Cui and Z. Wu and Y. Lu and X. Wei, and L. Dai, ``Near-field MIMO communications for 6G: Fundamentals, challenges, potentials, and future directions,''~\textit{IEEE Communications Magazine}, vol. 61, no. 1, pp. 40-46, Jan. 2023, doi: 10.1109/MCOM.004.2200136.

\bibitem{Liu1}
Y. Liu, Z. Wang, J. Xu, C. Ouyang, X. Mu, and R. Schober, ``Near-field communications: A tutorial review,''~\textit{IEEE Open Journal of the Communications Society}, vol. 4, pp. 1999-2049, 2023, doi: 10.1109/OJCOMS.2023.3305583.

\bibitem{Zeng}
H. Lu and Y. Zeng, ``Communicating with extremely large-scale array/surface: Unified modeling and performance analysis,''~\textit{IEEE Transactions on Wireless Communications}, vol. 21, no. 6, pp. 4039-4053, Jun. 2022, doi: 10.1109/TWC.2021.3126384.

\bibitem{R1-3}
Sadeghian, M., Pizzo, A., and Lozano, A. ``RIS in indoor environments: Benchmarking against ambient propagation,''~\textit{arXiv}, 2023, arXiv:2311.05266.

\bibitem{yufei-IoT}
Y. Zhao, Y. L. Guan, A. M. Ismail, G. Ju, D. Lin, Y. Lu, C. Yuen, ``Holographic-inspired meta-surfaces exploiting vortex beams for low-interference multipair IoT communications: From theory to prototype,''~\textit{IEEE Internet of Things Journal}, vol. 11, no. 7, pp. 12660-12675, Apr. 2024, doi: 10.1109/JIOT.2023.3334746.

\bibitem{NTP}
I. M. Salom,~\textit{et al.}, ``Implementation of NTP protocol in an isolated corporate network,''~\textit{2019 27th Telecommunications Forum (TELFOR)}, Belgrade, Serbia, 2019, pp. 1-4, doi: 10.1109/TELFOR48224.2019.8971119.

\bibitem{FEC}
E. A. Lee and D. G. Messerschmitt, Digital communication. Springer Science \& Business Media, 2012.

\bibitem{Yufei-TAP}
Y. Zhao, Z. Wang, Y. Lu, Y. L. Guan, ``Multimode OAM Convergent transmission with co-divergent angle tailored by Airy wavefront,'' ~\textit{IEEE Transactions on Antennas and Propagation}, vol. 71, no. 6, pp. 5256-5265, Jun. 2023, doi: 10.1109/TAP.2023.3263920.

\bibitem{Malu1}
C. Zhang, L. Ma, ``Millimetre wave with rotational orbital angular momentum,''~\textit{Scientific Reports}, vol. 6, p. 31921, 2016, doi: 10.1038/srep31921.

\bibitem{Malu2}
C. Zhang, L. Ma, ``Detecting the orbital angular momentum of electro-magnetic waves using virtual rotational antenna,''~\textit{Scientific Reports}, vol. 7, p. 4585, 2017, doi: 10.1038/s41598-017-04313-4.


\bibitem{ROAM}
J. Zhang, P. Li, R. C. C. Cheung, A. M. H. Wong, and J. Li, ``Generation of time-varying orbital angular momentum beams with space-time-coding digital metasurface,''~\textit{Advanced Photonics}, vol. 5, no. 3, p. 036001, Apr. 2023, doi: 10.1117/1.AP.5.3.036001.

\bibitem{ROAM2}
B. Liu, H. Chu, H. Giddens, R. Li, and Y. Hao, ``Experimental observation of linear and rotational Doppler shifts from several designer surfaces,''~\textit{Scientific Reports}, vol. 9, p. 8971, Jun. 2019, doi: 10.1038/s41598-019-45516-1. 

\bibitem{APMC-ge}
Y. Zhao, ~\textit{et al.}, ``OAM-based Reconfigurable Doppler Shifts Enable PAPR Reduction for Multi-carrier Doppler Diversity,''~\textit{2022 Asia-Pacific Microwave Conference (APMC)}, Yokohama, Japan, 2022, pp. 485-487, doi: 10.23919/APMC55665.2022.9999832.


\end{thebibliography}
\end{document}